\PassOptionsToPackage{unicode}{hyperref}
\PassOptionsToPackage{hyphens}{url}
\PassOptionsToPackage{dvipsnames,svgnames*,x11names*}{xcolor}
\documentclass[
  astrosymb, twocolumn, tighten, twocolappendix]{aastex631}
\usepackage{amsmath,amssymb}
\usepackage{lmodern}
\usepackage{ifxetex,ifluatex}
\ifnum 0\ifxetex 1\fi\ifluatex 1\fi=0 
  \usepackage[T1]{fontenc}
  \usepackage[utf8]{inputenc}
  \usepackage{textcomp} 
\else 
  \usepackage{unicode-math}
  \defaultfontfeatures{Scale=MatchLowercase}
  \defaultfontfeatures[\rmfamily]{Ligatures=TeX,Scale=1}
\fi
\IfFileExists{upquote.sty}{\usepackage{upquote}}{}
\IfFileExists{microtype.sty}{
  \usepackage[]{microtype}
  \UseMicrotypeSet[protrusion]{basicmath} 
}{}
\makeatletter
\@ifundefined{KOMAClassName}{
  \IfFileExists{parskip.sty}{%
    \usepackage{parskip}
  }{
    \setlength{\parindent}{0pt}
    \setlength{\parskip}{6pt plus 2pt minus 1pt}}
}{
  \KOMAoptions{parskip=half}}
\makeatother
\usepackage{xcolor}
\IfFileExists{xurl.sty}{\usepackage{xurl}}{} 
\IfFileExists{bookmark.sty}{\usepackage{bookmark}}{\usepackage{hyperref}}
\hypersetup{
  colorlinks=true,
  linkcolor=Maroon,
  filecolor=Maroon,
  citecolor=Blue,
  urlcolor=Blue,
  pdfcreator={LaTeX via pandoc}}
\usepackage{graphicx}
\makeatletter
\def\maxwidth{\ifdim\Gin@nat@width>\linewidth\linewidth\else\Gin@nat@width\fi}
\def\maxheight{\ifdim\Gin@nat@height>\textheight\textheight\else\Gin@nat@height\fi}
\makeatother
\setkeys{Gin}{width=\maxwidth,height=\maxheight,keepaspectratio}
\makeatletter
\def\fps@figure{htbp}
\makeatother
\usepackage{longtable,booktabs,array}
\usepackage{calc} 
\usepackage{etoolbox}
\makeatletter
\patchcmd\longtable{\par}{\if@noskipsec\mbox{}\fi\par}{}{}
\makeatother
\IfFileExists{footnotehyper.sty}{\usepackage{footnotehyper}}{\usepackage{footnote}}
\makesavenoteenv{longtable}
\usepackage[capitalize]{cleveref}
\usepackage{CJKutf8}
\usepackage{newunicodechar}

\newcommand{\Dnu}{\ensuremath{\Delta\nu}}
\newcommand{\numax}{\ensuremath{{\nu_\text{max}}}}

\defcitealias{ong_semianalytic_2020}{OB20}
\defcitealias{deheuvels_near_2017}{Dh17}

\newcommand\mesa{{\textsc{mesa}}}
\newcommand\gyre{{\textsc{gyre}}}

\newcommand{\chinesename}{{\begin{CJK}{UTF8}{gbsn}(王加冕)\end{CJK}}}

\DeclareRobustCommand{\okina}{%
  \raisebox{\dimexpr\fontcharht\font`A-\height}{%
    \scalebox{0.8}{`}%
  }%
}
\newunicodechar{ʻ}{\okina}

\newcommand{\annotate}[2]{\begin{tikzpicture}
    \node[anchor=south west,inner sep=0,align=center] (image) at (0,0) {
    #1
    };
    \begin{scope}[x={(image.south east)},y={(image.north west)}]
    #2
    \end{scope}
\end{tikzpicture}}
\def\bibitem{\vskip\bibskip\footnotesize\savebibitem}

\graphicspath{{./},{./figures/}}

\newcommand{\ind}[1]{_{\mathrm{#1}}}

\renewcommand{\edit}[2]{{\ifnum#1<9999%
#2%
\else%
\textbf{#2}%
\fi}}
\usepackage{mathptmx,txfonts,tikz,bm}
\ifluatex
  \usepackage{selnolig}  
\fi
\usepackage[]{natbib}
\defcitealias{ong_rotation_2022}{Paper I}
\newcommand\obb{{\citetalias{ong_rotation_2022}}}

\begin{document}

\shorttitle{$\zeta$, two ways}
\title{Mode Mixing and Rotational Splittings: II. Reconciling Different Approaches to Mode Coupling}
\correspondingauthor{Joel Ong}
\email{joelong@hawaii.edu}
\author[0000-0001-7664-648X]{J. M. Joel Ong \chinesename}
\affiliation{Institute for Astronomy, University of Hawaiʻi, 2680 Woodlawn Drive, Honolulu, HI 96822, USA}
\affiliation{Hubble Fellow}
\affiliation{Department of Astronomy, Yale University, 52 Hillhouse Ave., New Haven, CT 06511, USA}

\author[0000-0002-0833-7084]{Charlotte Gehan}
\affiliation{Max-Planck-Institut für Sonnensystemforschung, Justus-von-Liebig-Weg 3, 37077 Göttingen, Germany}
\affiliation{Instituto de Astrofísica e Ciências do Espaço, Universidade do Porto, CAUP, Rua das Estrelas, PT4150-762 Porto, Portugal}
\received{November 12, 2022}
\revised{January 31, 2023}
\accepted{February 23, 2023}
\submitjournal{\apj}
\shortauthors{Ong \& Gehan}
\def\sectionautorefname{Section}
\def\subsectionautorefname{Section}
\def\subsubsectionautorefname{Section}


\begin{abstract}
In the mixed-mode asteroseismology of subgiants and red giants, the coupling between the p- and g-mode cavities must be understood well in order to derive localised estimates of interior rotation from measurements of mode multiplet rotational splittings. There exist now two different descriptions of this coupling: one based on an asymptotic quantisation condition, and the other arising from coupling matrices associated with "acoustic molecular orbitals". We examine the analytic properties of both, and derive closed-form expressions for various quantities --- such as the period-stretching function $\tau$ --- which previously had to be solved for numerically. Using these, we reconcile both formulations for the first time, deriving relations by which quantities in each formulation may be translated to and interpreted within the other. This yields an information criterion for whether a given configuration of mixed modes meaningfully constrains the parameters of the asymptotic construction, which is likely not satisfied by the majority of first-ascent red giant stars in our observational sample. Since this construction has been already used to make rotational measurements of such red giants, we examine --- through a hare-and-hounds exercise --- whether, and how, such limitations affect existing measurements. While averaged estimates of core rotation seem fairly robust, template-matching using the asymptotic construction has difficulty reliably assigning rotational splittings to individual multiplets, or estimating mixing fractions $\zeta$ of the most p-dominated mixed modes, where such estimates are most needed. We finally discuss implications for extending the two-zone model of radial differential rotation, e.g. via rotational inversions, with these methods.
\keywords{Asteroseismology (73), Red giant stars (1372), Stellar oscillations (1617), Computational methods (1965), Theoretical techniques (2093)}
\end{abstract}

\hypertarget{introduction}{%
\section{Introduction and Motivation}\label{introduction}}

The asteroseismology of solar-like oscillators in the age of high-precision space photometry missions --- such as CoRoT, Kepler and TESS --- increasingly must contend with modes of mixed character, associated with solar-like oscillators which have evolved off the main sequence. In particular, whereas stars like the Sun can be seen to support pressure waves, whose p-mode frequencies approximately satisfy
\begin{equation}
    \nu_{n_plm} \sim \Dnu\left(n_p + {l \over 2} + \epsilon_{p, n_plm}\right),\label{eq:p}
\end{equation}
(where \(\Dnu\) is a characteristic frequency spacing \edit1{called the large frequency separation}, and \(n_p\), \(l\), and \(m\) are integer quantum numbers), the observed waves of these these mixed-mode oscillators also propagate as gravity waves in their interiors, trapped within their radiative cores. Notionally, these g-mode frequencies should satisfy
\begin{equation}
    {1 \over \nu_{n_glm}} \sim \Delta\Pi_l\left(n_g + \epsilon_{g, n_glm}\right),\label{eq:g}
\end{equation}
where \(\Delta\Pi_l\) is a characteristic period spacing. In practice, however, those modes which are observed possess mixed character, propagating buoyantly in the core and acoustically in the envelope. This has proven to be a mixed blessing: on one hand, these mixed modes yield highly sensitive observational probes into otherwise inaccessible stellar interiors. On the other, the coupling between the two mode cavities disrupts the simple comb-like structure of the mode frequencies specified by \cref{eq:p,eq:g}, which may no longer be directly relied upon. The interpretation of these mixed-mode frequencies must thus differ significantly from that of pure p- or g-modes considered in isolation.

One crucial diagnostic quantity for these mixed modes is their mixing fraction, typically denoted as \(\zeta\), which describes how much a mode under consideration resembles a pure g-mode (where \(\zeta \to 1\)) or a pure p-mode (\(\zeta \to 0\)). When the radiative core is much more compact than the star, this quantity may be found directly from the mixed-mode eigenfunctions as a ratio of partial mode inertiae:
\begin{equation}
    \zeta_{nlm} \sim { \int_\text{only radiative core} |\bm{\xi}_{nlm}|^2 \mathrm d m \over \int_\text{whole star} |\bm{\xi}_{nlm}|^2 \mathrm d m},\label{eq:ratios}
\end{equation}
where $m(r) = \int_0^r 4\pi r'^2 \rho \mathrm d r'$ is the mass coordinate, and $\bm{\xi}$ is the Lagrangian displacement eigenfunction. In practice, the structure of these stars is not directly accessible, and so these mixing fractions must instead be inferred. The inferred mixing fractions are in turn used to interpret other seismic measurements. For instance, when these stars rotate, their dipole modes exhibit rotational splitting into triplets. To leading order in perturbation theory, the frequency widths of these triplets may be approximated \edit1{using} a simplified two-zone model of radial differential rotation \edit1{(with only globally averaged core and envelope rotation rates $\Omega_\text{core}$ and $\Omega_\text{env}$)} as \cite[e.g.][]{goupil_seismic_2013}
\begin{equation}
  \delta\omega_{nl} \sim \omega_\text{rot} - \omega_\text{nonrot} \sim m \left[ \beta_g \zeta \Omega_\text{core} + \beta_p (1 - \zeta) \Omega_\text{env}\right].\label{eq:rotation}
\end{equation}
Here $\beta_p$ and $\beta_g$ are sensitivity constants associated with pure p- and g-modes.
This relation is best satisfied when the rotational splitting is small --- i.e. in less evolved subgiant or red giant stars, or more generally for the most p-dominated mixed modes. The conversion between the observed rotational splittings and the interpreted core rotation rates in these cases thus depends critically on the inferred values of \(\zeta\) \citep[e.g.][]{deheuvels_near_2017,deheuvels_seismic_2020,triana_2017}. As such, it is in these situations where accuracy in inferring the mixing fractions is paramount.

However, \cref{eq:rotation} requires that the rotational splittings be linear, which is not generally the case; in more evolved red giants a highly nonlinear phenomenology of dense avoided crossings emerges \citep[e.g.][]{ouazzani_rotation_2013}, rendering first-order perturbation analysis no longer \edit1{generally} tenable. To treat these avoided crossings, more sophisticated techniques are required. There are now two distinct analytic treatments of this mode mixing by which such inferences may be made. Of older vintage, the asymptotic analysis of mixed-mode oscillators in the JWKB approximation \citep[e.g.][]{shibahashi_modal_1979,unno_nonradial_1989, takata_asymptotic_2016} relates mixed-mode wavefunctions to stellar structure in a local fashion, with the mixed-mode frequencies themselves emerging as roots of some characteristic eigenvalue equation. More recently, an algebraic approach has also been developed \citep[e.g.][]{deheuvels_insights_2010, ong_semianalytic_2020, ong_surface_1,ong_rotation_2022}, where the pulsation problem is first modified to yield a nonorthogonal ``decoupled'' set of pure-p and pure-g-mode basis functions; the mixed-mode frequencies then emerge as the roots of the characteristic equation associated with a generalised Hermitian matrix eigenvalue problem, as represented in this nonorthogonal basis. Given such a set of basis functions, inferences of the mixing fraction $\zeta$ for each mode permit an effective p- and g-mode splitting to be assigned to that mode, potentially then allowing for generalisations to the two-zone model of differential rotation \cite[i.e. rotational inversion, as in e.g.][]{schunker_inversion_2016a,schunker_inversion_2016b} to be prosecuted for these evolved red giants \citep[as attempted in e.g.][]{di_mauro_2016, di_mauro_2018, fellay_asteroseismology_2021}.

Associated with these two analytic constructions are two distinct approaches to operationalising them for use with actual observational data. Thus far, the matrix construction has only been used in the context of direct forward modelling against grids of evolutionary structural models \citep[e.g.][]{ong_surface_2}. Such forward modelling is very expensive, but at the end of the day produces guesses at the structure of the star by which \cref{eq:ratios} may be evaluated directly, obviating the need for further analytic approximations. By contrast, the analytic properties of the asymptotic approach, along with a generative parameterisation for the pure p- and pure g-mode frequencies, have been put together to yield a more flexible diagnostic construction --- that of `stretched \'echelle diagrams' \citep{mosser_period_2015, gehan_core_2018, mosser_period_2018} --- proposed for direct use with actual power spectra, and unencumbered by any requirement for stellar modelling (at the expense of needing to fit free parameters).

This generative property has rendered the asymptotic construction particularly compelling for use on highly evolved red giant stars, where JWKB analysis is assumed to hold good, and where in any case the evolutionary and algebraic calculations associated with the matrix methods become increasingly expensive. At the same time, however, the asymptotic approach is also not free of defects. For one, its free parameters are strictly phenomenological, and no complete interpretation of them has been supplied. Observationally, the deviations between its predictions and actual mixed-mode frequencies are largest in the neighbourhood of p-dominated mixed modes \citep[e.g.~Fig. 3 of][]{mosser_period_2018} --- precisely where its correctness is most necessary, since it is only these modes which are observed with any appreciable amplitude. \edit1{Conventionally, this is regarded as a consequence of some approximations made in} its underlying \edit1{analytic} construction. \edit1{Since these outstanding issues afflict the p-dominated mixed modes most significantly}, existing users of the stretched \'echelle diagram construction have elected instead to omit the p-dominated mixed modes entirely from their analysis.

\edit1{In this series of papers, we attempt to extend the two-zone model of differential rotation in the presence of these avoided crossings, with the ultimate goal of performing rotational inversions. As discussed in \citet[hereafter \obb]{ong_rotation_2022}, the notional pure g-modes in red giants are deep into the asymptotic regime, and thus yield little information beyond a single averaged core rotation rate, even collectively; accordingly, in this work we place a particular focus on the p-dominated mixed modes. In \autoref{analytic-developments}, we therefore critically reexamine the asymptotic construction, and analytically justify its applicability to these modes specifically. Since the matrix-based analytic approach discussed in \obb{} is computationally expensive to operationalise, we also examine in this section the prospects for applying observational results derived from the asymptotic procedure, which is significantly cheaper, within the matrix construction. Even within our fortified asymptotic construction, however, our discussion also indicates that the stretching function of a given configuration of mixed modes may be meaningfully constrained only if a sampling-based information criterion is satisfied. In \autoref{numerical-results}, we find that this criterion is not satisfied for much of the existing observational sample. Given that such stretching functions have already extensively been used to characterise stellar properties, we explore the implications of this result, via a hare-and-hounds exercise against one existing observational pipeline for rotational measurements. Finally, we summarise our findings, and discuss future directions, in \autoref{sec:discussion}}.

\hypertarget{analytic-developments}{%
\section{Analytic Developments}\label{analytic-developments}}

\begin{figure}[htbp]
    \centering
    \begin{tikzpicture}[xscale=.8, yscale=.8]
\draw[dashed, gray] (-4,-.25) grid (4,0);
\draw[->, line width=1 pt] (-5, 0) -- (5, 0);
\foreach \x in {-4, -3, -2, -1, 0, 1, 2, 3, 4}{
  \node at (\x, .5){\small $\nu_{g,\x}$};
  \draw[red, line width=1 pt] (\x, 0) -- (\x, .35);
}
\foreach \x in {-4, -3, -2, -1, 1, 2, 3, 4}{
  \node at (\x, -.5){\small $\nu_{\x}$};
  \draw[red, line width=1 pt] (\x + 0.1 / \x - 0.01, 0) -- (\x + 0.1 / \x - 0.01, -.35);
}
\node at (-.25, 1){\small $\nu_p$};
\draw[blue, line width=1 pt](-.25, 0) -- (-.25, .85);
\node at (-.5, -1){\small $\nu_-$};
\draw[purple, line width=1 pt](-.4, 0) -- (-.4, -.85);
\node at (.2, -1){\small $\nu_+$};
\draw[purple, line width=1 pt](.16, 0) -- (.16, -.85);
\node at (-5.3, 0){$\cdots$};
\end{tikzpicture}
    \caption{Configuration of pure and mixed modes. Above the frequency axis, we mark out the notional locations of the pure g-modes, with frequencies $\nu_{g,i}$, in the vicinity of a pure p-mode, with frequency $\nu_p$. Below the frequency axis, we mark out the locations of the resulting mixed modes, both far away from resonance (indicated with the integer indices as $\nu_i$), and the two closest to resonance (labelled as $\nu_\pm$).}
    \label{fig:axis}
\end{figure}
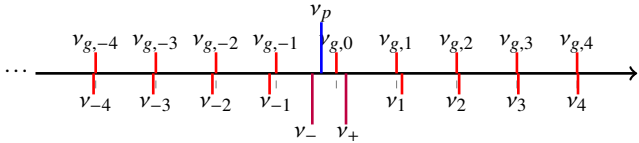

We consider mixed modes arising from a dense set of pure g-modes coupling to a sparse set of p-modes. We will limit our attention to the g-modes closest to a particular p-mode with frequency \(\nu_p\). We label the frequency of the g-mode closest to this p-mode as \(\nu_{g,0}\), and those of the adjacent g-modes with integer indices \(i\) as \(\nu_{g,i}\). The coupling between these pure modes gives rise to a pair of mixed modes \(\nu_{\pm}\) close to the on-resonance pair \(\nu_p\) and \(\nu_{g,0}\), as well as further mixed modes \(\nu_i\). Critically, the frequency of the p-mode \(\nu_p\) is completely independent of the properties of the g-modes \(\nu_{g, i}\). We illustrate this configuration in \cref{fig:axis}.

In the presence of rotation, each mode splits into a $(2\ell + 1)$-tuplet. The interactions between rotation and mixed-mode coupling are such that for fast enough rotation, the linear expression \cref{eq:rotation} may cease to hold for the mixed modes. However, in \obb{}, we also demonstrate that the rotational perturbations to the underlying p- and g-modes may still be treated as linear in these cases. Moreover, the interactions between mode coupling and rotation are such that, to a good approximation, the components of the same $m$ from each multiplet may be considered to couple to each other independently of the other $m$. We thus restrict our following discussion to modes of the same $m$, on the understanding that the same calculations must be performed repeatedly for each $m$ to recover the full set of observed modes.


We now derive the basic properties of the mixed-mode frequencies with respect to both of the existing analytical frameworks we have previously described.

\hypertarget{asymptotic-construction}{%
\subsection{Asymptotic Construction}\label{asymptotic-construction}}

In the JWKB construction \edit2{of} \citep{unno_nonradial_1989}, the frequency eigenvalues emerge from an eigenvalue equation \edit1{associated with a characteristic function}
\begin{equation}
    F(\nu) = \tan \theta_p(\nu) \cot \theta_g(\nu) - q(\nu) \text{ such that } F(\nu_i) = 0,\label{eq:eigjwkb}
\end{equation}
where \(\theta_p\), \(\theta_g\), and \(q\) are frequency-dependent phase or tunnelling integrals over the stellar structure with respect to an appropriate JWKB dispersion relation. Note that the characteristic function \(F(\nu)\) is not generally 0 away from the eigenvalues. \citet{mosser_period_2015}, and subsequent works, neglect the frequency dependence of \(q\), and replace these phase integrals with expressions involving the underlying p- and g-modes as
\begin{equation}
    \theta_p \mapsto \Theta_p \equiv \pi{\nu - \nu_p \over \Dnu};\ \ \  \theta_g \mapsto \Theta_g \equiv \pi{{1 \over \nu_g} - {1 \over \nu} \over \Delta\Pi_l}.\label{eq:bigtheta}
\end{equation}

These expressions are chosen so that \(\Theta_p \to n \pi\) as \(\nu \to \nu_p\) for some integer \(n\), and likewise for \(\Theta_g\) and \(\nu_g\). \edit1{Moreover, as multiplication by a constant does not change the roots of the characteristic function, we have chosen an overall sign for our description of it to yield a corresponding positive sign of $q$ in many of our subsequent expressions}. Before we proceed, we must now break with the existing analysis in pointing out that this identification of \(\theta_g\) with \(\Theta_g\) results in inconsistent behaviour. Specifically, the frequencies of the most g-dominated mixed modes should be best described by those of the underlying g-modes: that is, \(\nu_i \to \nu_{g, i}\) far from resonance with a p-mode \citep[e.g.][]{jcd_kepler_2012}. For these modes, \(\tan \Theta_p\) takes on very large absolute values, as \(\Theta_p(\nu_g) \to \pi \left(n + {1 \over 2}\right)\). Accordingly, the eigenvalues of these g-dominated modes must satisfy \(\cot \theta_g \to 0\) as \(\nu \to \nu_g\). But \(\Theta_g \to n\pi\) as \(\nu \to\nu_g\), so \(\cot \Theta_g = \tan\left({\pi \over 2} - \Theta_g\right)\) is also singular as \(\nu \to \nu_g\), and therefore cannot satisfy \cref{eq:eigjwkb} --- a contradiction. This inconsistency is resolved by identifying \(\theta_g\) with \({\pi \over 2} - \Theta_g\) instead, resulting in a phase offset of \({\pi \over 2}\) relative to \edit1{almost} all literature expressions involving \(\Theta_g\) in our subsequent discussion. \edit2{This phase offset can be motivated analytically, relative to the formulation of \cite{unno_nonradial_1989}, by either working outside of the Cowling approximation \cite[as in][]{takata_asymptotic_2016}, or by using the Lagrangian pressure perturbation (rather than displacement) as the dynamical variable \cite[as in][]{gough_linear_1993}}. \cite{lindsay_overshoot_2022} have demonstrated that such a phase offset is indeed necessary for agreement between \edit1{values for $\epsilon_g$ from the sample of \cite{mosser_period_2018}}, and \edit1{from} the most g-dominated mixed modes that emerge from explicit numerical frequency calculations made without the JWKB approximation. \edit1{Such a correction also resolves the implicit inconsistency in Fig. 12 of \cite{mosser_period_2018}, where the Cowling-approximation asymptotic value of \cite{provost_asymptotic_1986} --- which does not suffer from this issue --- appears not to exhibit the predicted offset of 1/2 caused by the Cowling approximation, relative to the values of \cite{takata_asymptotic_2016} --- made without the Cowling approximation, but also suffering from this issue. While the expressions of \cite{mosser_period_2018} may be alternatively (and charitably) interpreted as simply encoding a different convention for $\epsilon_g$ in mixed modes vs. pure g-modes, it would still be more desirable for us to, in the first place, choose a convention in which the values of $\epsilon_g$ from the pure g-mode frequencies were consistent with those computed directly from the most g-dominated mixed modes, as they should approach each other far from resonance (cf. \cref{fig:axis}).} We thus rewrite the characteristic function \cref{eq:eigjwkb} and the associated eigenvalue equation as
\begin{equation}
    F(\nu) \approx \tan \Theta_p(\nu) \tan \Theta_g(\nu) - q(\nu) \text{ such that } F(\nu_i)= 0.\label{eq:eig}
\end{equation}
The definitions of \(\Theta_p\) and \(\Theta_g\) given in \cref{eq:bigtheta} have so far been interpreted in a local sense as functions of a single variable \(\nu\), where \(\nu_p\) and \(\nu_g\) are taken to be specified by the nearest p- and g-mode frequencies to \(\nu\)\edit1{, and held fixed}. However, we note that these definitions may be generalised to yield \emph{bona fide} multivariable functions in a self-consistent fashion. In particular, \cref{eq:p,eq:g} suggest definitions of angular quantities as \citep{mosser_probing_2012}
\begin{equation}
    \theta_p(\nu) = \pi \left({\nu \over \Dnu} - {l \over 2} - \epsilon_p(\nu)\right);\ \ \ \theta_g(\nu) = \pi \left({1 \over \nu \Delta\Pi_l} - \epsilon_g(\nu) + {1\over 2}\right),
\end{equation}
yielding eigenvalues where \(\theta_p(\nu_p) = n_p\pi\) and \(\theta_g(\nu_g) = \left(n_g + {1 \over 2}\right)\pi\) at integers \(n_p\) and \(n_g\). Accordingly, \cref{eq:bigtheta,eq:eig} can be seen to be consistent with the definitions
\begin{equation}
    \Theta_p(\nu, \nu_p) = \theta_p(\nu) - \theta_p(\nu_p);\ \ \ \Theta_g(\nu, \nu_g) = \theta_g(\nu_g) - \theta_g(\nu).
\end{equation}
With these modified definitions, the characteristic function may likewise be treated as a multivariable function in \(\nu\), \(\nu_g\), and \(\nu_p\). \cref{eq:eig} then specifies mixed-mode eigenvalues when \(F(\nu, \nu_g, \nu_p) = 0\) with \(\nu_p\) and \(\nu_g\) set to any combination of p- and g-mode eigenvalues, respectively. In our subsequent discussion, we hold \(\nu_p\) fixed at just such a p-mode eigenvalue (it does not matter which), but permit \(\nu_g\) to vary freely.

With this machinery in hand, we may now critically examine the assumptions going into the usual construction of stretched \'echelle diagrams. \edit1{We first reframe the standard derivation of the asymptotic parameterisation, with respect to our analytic continuation of $\nu$ and $\nu_g$ as independent continuous variables}. Far from resonance, the frequency of a given mixed mode \(\nu_i\) can be well approximated as a function of the nearest associated g-mode frequency \(\nu_{g,i}\) (with reference to \cref{fig:axis}). Since \(F(\nu_i, \nu_{g,i}) = 0\) for all mixed modes \(\nu_i\) and g-modes \(\nu_{g, i}\), the local spacing between mixed modes can be shown to approximately satisfy \citep{jcd_kepler_2012}
\begin{equation}
\begin{aligned}
{\delta P \over \Delta\Pi_l} &= {1/\nu_{i+1} - 1/\nu_{i} \over 1/\nu_{g,i+1} - 1/\nu_{g,i}} \sim {\nu_{g, i}^2 \over \nu_i^2} {\nu_{i+1} - \nu_i \over \nu_{g,i+1} - \nu_{g,i}} \\ &\sim  {\nu_{g}^2 \over \nu^2}\left.{\partial \nu \over \partial \nu_g}\right|_F = - {\nu_{g}^2 \over \nu^2} {{\partial F / \partial \nu_g}|_\nu \over {\partial F / \partial \nu}|_{\nu_g}} \\&= \left[1 + {\nu^2 \Delta\Pi_l \over \Dnu}{\tan \Theta_g(\nu, \nu_g) \cos^2 \Theta_g(\nu, \nu_g) \over \tan \Theta_p(\nu) \cos^2 \Theta_p(\nu)}\right]^{-1} \equiv \zeta(\nu, \nu_g).\label{eq:zeta1}
\end{aligned}
\end{equation}
where we \edit1{have used the identity\(\left.\partial x \over \partial y\right|_z\left.\partial y \over \partial z\right|_x\left.\partial z \over \partial x\right|_y = -1\), and} neglect the frequency dependence of \(q\), and of the g-mode phase function \(\epsilon_g\). The dependences on \(\Theta_g\), and so \(\nu_g\), may be further eliminated when this is evaluated at eigenvalues \(\nu_i\), or more generally on level sets where \(F(\nu, \nu_g) = 0\) (via \cref{eq:eig}), as
\begin{equation}
    \zeta(\nu_i, \nu_{g, i}) = \zeta_p(\nu_i) \equiv \left[1 + {\Delta\Pi_l \over \Dnu}{\nu^2 \over q \cos^2 \Theta_p(\nu_i) + {1\over q}\sin^2 \Theta_p(\nu_i)}\right]^{-1},\label{eq:zeta2}
\end{equation}
which, fortuitously, does not suffer from the systematic offset of \(\pi \over 2\) which we have discussed above. We are fortunate indeed that it is this expression that has found widespread practical application. Supposing that this function \(\zeta_p\) were directly accessible, \citet{goupil_seismic_2013} find that in the JWKB approximation, its values, evaluated at the mixed mode frequencies \(\nu_i\), also approximate the mixing fractions specified by inertia ratios, \cref{eq:ratios}.

\edit1{We now consider the existing construction of stretched period-\'echelle diagrams with respect to this analytical framework. Having followed through an analogous derivation of \cref{eq:zeta1} \cite[also following][]{jcd_kepler_2012}, \cite{mosser_period_2015} then observed that the finite-difference approximation,}
\begin{equation}
{\delta P \over \Delta\Pi_l} \sim \zeta(\nu, \nu_g), \label{eq:requiredproperty}
\end{equation}
\edit1{may be replaced by ansatz with an equality,}
\begin{equation}
    {\mathrm{d} \tau \over \mathrm{d} P} = {1 \over \zeta(1/P, \nu_g)},\label{eq:ode}
\end{equation}
\edit1{such that the coordinate $\tau$ that is so defined accumulates increments of $\Delta\Pi_l$ when $P$ is integrated over the intervals between adjacent mixed-mode eigenvalues $1/\nu_i$, by construction. In their treatment, the g-mode frequency $\nu_g$ is held fixed, yielding highly oscillatory behaviour when the integrand is evaluated as a continuous function of $\nu$ (cf. their fig. 2). We note that this is inconsistent with the definition of $\zeta$ as a directional derivative along connected level sets of $F$, as can be seen from our \cref{eq:zeta1}. In a bid to eliminate this oscillatory behaviour (and better condition their numerical integration), they then noted, as above, that evaluating $\zeta(\nu_i, \nu_g)$ only at mixed-mode eigenvalues $\nu_i$ (i.e. on a disconnected level set of $F = 0$ with constant $\nu_g$) yielded values of what appeared to be a far simpler function ($\zeta_p$; \cref{eq:zeta2}), which also produced the desired properties upon integration when inserted into \cref{eq:ode}. Thus, despite their general inequivalence on lines of constant $\nu_g$, $\zeta$ and $\zeta_p$ were identified with each other for the purposes of integrating \cref{eq:ode}: first by interpolation in \cite{mosser_period_2015}, and then explicitly in \cite{mosser_period_2018}.}

\edit1{We now proceed with an alternative derivation that does not suffer from these issues. First, we note that, upon analytic continuation, there} is some redundancy in the definition of this \(\tau\) coordinate, via \cref{eq:zeta1,eq:ode}, which may not be immediately obvious. Since we are entitled to treat \(F(\nu, \nu_g)\) formally as a function of two variables, we have seen that \(\zeta\) and \(\zeta_p\) are equivalent only on level sets of \(F\) on the \(\nu\)-\(\nu_g\) plane where \(F(\nu, \nu_g) = 0\). This being the case, \cref{eq:zeta1} defines these level sets near the line of equality (where \(\nu = \nu_g\)) via an ordinary differential equation of the form \(\mathrm d \nu / \mathrm d \nu_g = f(\nu, \nu_g)\).
Explicitly, we may parameterise these level sets in differential form as
\begin{equation}
  \mathrm{d}F = 0 = \left.\partial F \over \partial \nu_g\right|_\nu\mathrm d \nu_g + \left.\partial F \over \partial \nu\right|_{\nu_g}\mathrm d \nu \implies {\mathrm d \nu \over \mathrm d \nu_g} = - {{\partial F / \partial \nu_g}|_\nu \over {\partial F / \partial \nu}|_{\nu_g}} = {\nu^2 \over \nu_g^2} \zeta(\nu, \nu_g). \label{eq:integralcurve}
\end{equation}
\edit1{\cref{eq:ode} is then recovered upon the change of variables to $\tau \equiv 1/\nu_g$ and $P = 1/\nu$.} The prescription of \citet{mosser_period_2015} is \edit1{thus equivalent to producing integral curves of \cref{eq:integralcurve} in the \(\nu\)-\(\nu_g\) plane from an initial value problem (IVP), which can be seen to yield connected level sets of $F = 0$ only when passing through some fiducial point \((\nu_i, \nu_{g,i})\) near the line of equality. Only on such an integral curve, we are now correctly entitled to replace $\zeta$ with $\zeta_p$ when performing this integration. It is extremely fortunate that \cite{mosser_period_2015} achieve the correct result through an even number of cancelling errors, so that observational results relying upon this construction remain valid, at least conceptually.}

\begin{figure}
\centering
\includegraphics{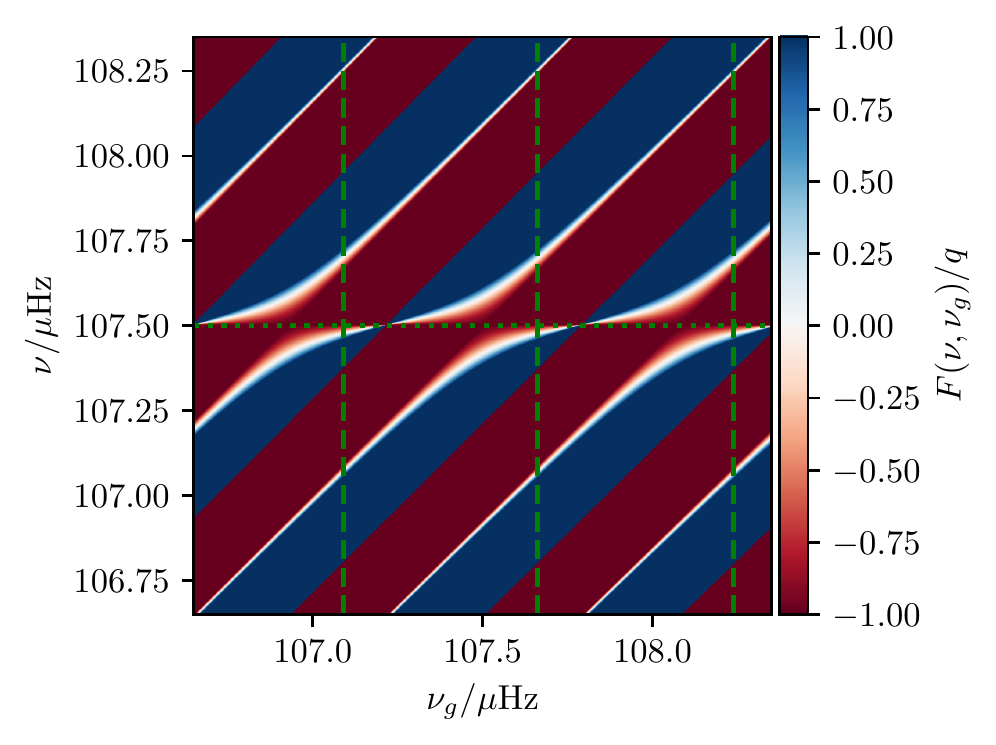}
\caption{Values of the characteristic function \(F\) on the \(\nu-\nu_g\) plane near the line of equality (the diagonal of the figure), for a simplified \edit1{illustrative} configuration of mixed modes. The colour scale is deliberately saturated so as to accentuate the loci of level sets where \(F = 0\), corresponding to the white regions. The frequencies of the g-mode eigenvalues \(\nu_{g, i}\) are shown with the green dashed lines, while the nearest p-mode frequency \(\nu_p\) is shown with a green dotted line. \label{fig:levelsets}}
\end{figure}

\edit1{A discrete family of such connected level sets exists on the \(\nu\)-\(\nu_g\) plane.} To build graphical intuition for this construction, we show in \cref{fig:levelsets} an example of the values of \(F\) one might obtain on the \(\nu\)-\(\nu_g\) plane near the line of equality, emphasising the loci of these level sets by saturating our colour map, for \edit1{an illustrative configuration} of these mixed modes (generated with constant values of \(q\), \(\epsilon_p\), and \(\epsilon_g\)). Each of the white loci are a different level set, \(\nu = f_n(\nu_g)\). In this restricted scenario, the level sets are periodic as \(f_n(\nu_g) = f_{n-k}\left([1/\nu_g + k\Delta\Pi_l]^{-1}\right)\), although this is not generally true. Morphologically, these level sets resemble diagonal lines parallel to the line of equality, undergoing avoided crossings with horizontal lines \(\nu = \nu_{p}\) at the pure p-mode eigenvalues (dotted line).

Mixed-mode frequencies may be read directly off \cref{fig:levelsets} by searching for intersections between these level sets (white loci) with vertical lines specified by the underlying g-mode frequencies (dashed lines) --- i.e.~by evaluating \(f_n(\nu_g)\) at g-mode eigenvalues \(\nu_{g,i}\). Per \cref{eq:zeta1}, \(\zeta\) may also be read off the diagram as the gradients of these level sets at such points of intersection which are nearest to the line of equality: that is to say, \(\zeta_i = \zeta_p(\nu_i) = {\partial f_n \over \partial \nu_g}(\nu_g = \nu_{g, i})\), with \(n\) chosen to yield the level set nearest the line of equality.

\edit1{Having demonstrated that the integral curves generated by \cref{eq:ode,eq:integralcurve} are indeed level sets of $F = 0$, however, we can see immediately that} integrating \cref{eq:ode} to recover a level set is actually unnecessary, as \cref{eq:eig} already specifies these level sets in closed form. In particular we have
\begin{equation}
\begin{aligned}
    {1 \over \nu_g} = \tau(\nu) &\equiv \theta_g^{-1} \left(\theta_g(\nu) + \arctan \left[q \over \tan \Theta_p(\nu)\right]\right) \\ & \sim {1 \over \nu} + {\Delta\Pi_l \over \pi}\arctan\left(q \over \tan \Theta_p(\nu)\right) - \left(\nu - \nu_g\right){\partial \epsilon_g \over \partial \nu} + \ldots\label{eq:levelset}
\end{aligned}
\end{equation}
\edit1{This is most useful near the line of equality where \(\nu \sim \nu_g\), so that any frequency dependence of \(\epsilon_g\) may be neglected. We note that similar expressions were derived in \cite{mosser_period_2018,pincon_multicavity_2022}, but were not recognised to be a closed-form expression for the stretching function $\tau$, owing to the previously discussed phase offset of $\pi / 2$. Since such an expression exists, however, we recommend that it be used in future applications, in preference over numerical integration of \cref{eq:ode}. Because this expression is exact, and does not depend on the finite-difference approximation made by going from \cref{eq:requiredproperty} to \cref{eq:ode}, it is in principle also applicable to even the most p-dominated mixed modes. As such, the poor performance of the stretched-\'echelle-diagram construction for these modes in particular \cite[e.g. Fig. 3 of][]{mosser_period_2018} cannot be attributed to the poor quality of such a finite-difference approximation, as is customary in existing work using it. Instead, this must be the result of either the stretching function potentially not having been correctly evaluated numerically by integration (which we explore in more detail in \autoref{sec:integration}), or of an incorrect stretching function, correctly evaluated numerically, being used to describe a given set of mixed modes.}

\subsubsection{Constraining the stretching function}
\label{sec:overestimate}

\edit1{So far, our discussion has been generative, in the sense that we have described curves on the $\nu$-$\nu_g$ plane generated by $\zeta_p$, assumed to be known perfectly. However, in practice, neither the pure p- or g-mode eigenvalues, nor $\zeta$, are known: all of them have to be inferred solely from the mixed-moded eigenvalues $\nu_i$. In principle, this is an underdetermined problem, as the number of p- and g-mode eigenvalues is unchanged when they are coupled to yield mixed modes, but extra properties of the coupling (i.e. the coupling strength $q$) must additionally be inferred. For most red giants, however, a single effective value of $\Delta\Pi_l$ and $\epsilon_g$ may be assumed, considerably reducing the number of unknowns. Many prescriptions exist by which these properties may be derived, including autocorrelation-based techniques \citep[e.g.][]{vrard_period_2016,mosser_period_2017}, parametric fits to the mixed-mode frequencies \citep[e.g][]{jiang_variations_2020,jiang_evolution_2022,li_magnetic_2022,deheuvels_strong_2023}, or various constructions involving the mixed-mode period differences \cite[e.g.][]{cunha_analytical_2019,dreau_dipolar_2020,farnir_eggmimosa_2021}. While much of the existing literature has focused on accurate determination of $\Delta\Pi_l$, we focus here on constraining the p-mode frequencies $\{\nu_{p, i}\}$ and the coupling strength $q$ instead.}

\edit1{The p-mode frequencies are notionally located at minima of the mixed-mode period differences, although procedures for recovering them built upon this \citep[even incorporating frequency-dependent $q$ and more sophisticated fitting techniques, e.g.][]{dreau_dipolar_2020} yield results that systematically deviate from the pure p-modes derived from stellar models (Saunders et al., accepted to ApJ). Our construction provides a further constraint on their values. For the two most p-dominated mixed modes closest to resonance (i.e.~\(\nu_\pm\) of \cref{fig:axis}), \(\zeta_\pm\) can be seen to be the derivatives, evaluated at the same value of \(\nu_g = \nu_{g,0}\), of two} different level sets adjacent to the line of equality. Conversely, \edit1{perfect knowledge of $\tau$ would yield} $\tau(1/\nu_+) = \tau(1/\nu_-) = 1/\nu_{g, 0}$. \edit1{In a rotationally split multiplet, this criterion must be satisfied separately for each multiplet component indexed by $m$.  In principle, \cref{fig:levelsets} indicates that this is exact, even when the actual period difference between pure g-modes of adjacent order is not a constant $\Delta\Pi_l$ (e.g. due to curvature from buoyancy glitches, rotational splitting, or magnetic fields). However, an overall difference of some integer multiple of $\Delta\Pi_l$ may be introduced from alternative choices of branch for the arctan function when \cref{eq:levelset} is evaluated numerically. We propose this property to be} used as a \edit1{diagnostic} check of correctness for whether $\tau$ has been properly constructed to match \edit1{$\{\nu_{p, i}\}$ for} any given set of mixed modes.

\begin{figure}
\centering
\includegraphics{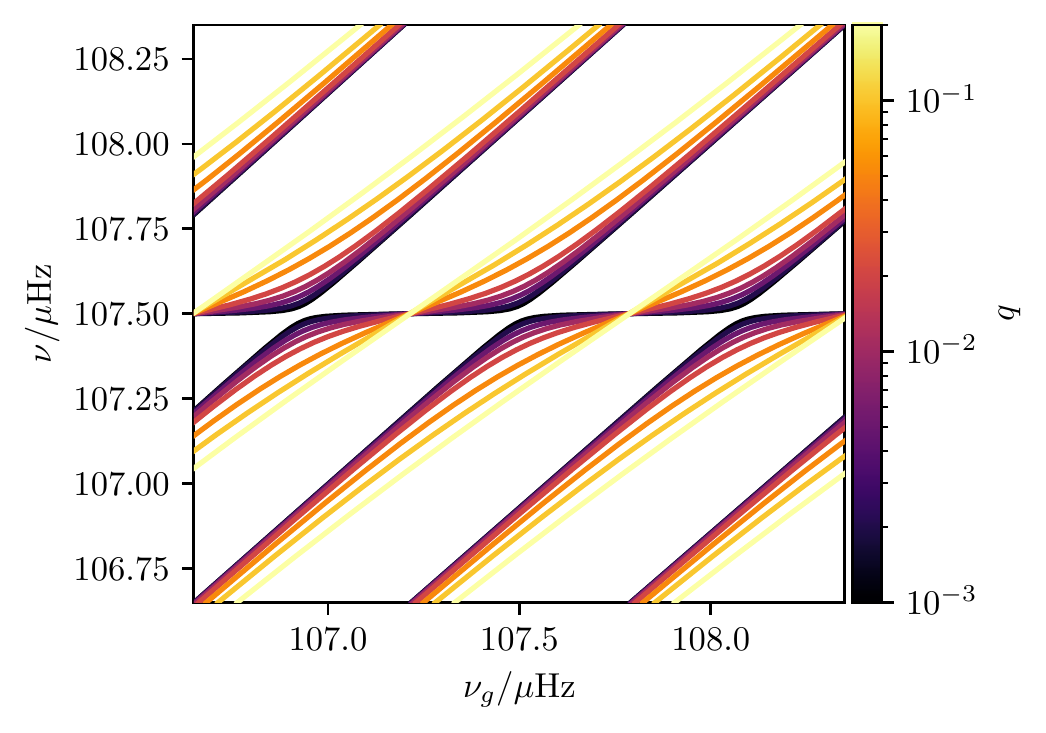}
\caption{Avoided crossings between adjacent level sets of \(F = 0\) in the \(\nu\)-\(\nu_g\) plane for the same configuration as in \cref{fig:levelsets}, coloured by different values of the coupling strength \(q\). \edit1{Eight values of $q$ are shown, at equal logarithmic spacing going from 0.001 to 0.2, inclusive.} Within this simplified configuration (in particular, constant \(q\) with no frequency dependence) these are also \edit1{the} contour lines of \cref{fig:levelsets}. \label{fig:q}}
\end{figure}

\edit1{We now examine} how observational data may also fail to constrain \edit1{\(q\), and thus \(\zeta_p\). While we will focus on the direct fitting of period differences in our subsequent discussion (as it is more analytically tractable than assessing the properties of a least-squares fit or periodogram), we observe that constraining power on $q$ is always, in these red giants, obtained from the mixed modes closest to the underlying p-mode. Moreover, taking period differences does not substantively decrease the amount of information available from the mixed modes, so these considerations should be generally applicable irrespective of actual observational practice. To simplify matters, we will assume a priori that the pure p-mode frequencies have somehow already been perfectly determined}. By inspection, estimating \(\zeta(\nu)\) directly using the period differences, as done in \cref{eq:requiredproperty}, is equivalent to evaluating \(f_n'(\nu_g)\) using a second-order finite-difference scheme, with step size \(h = \nu_g^2 \Delta\Pi_l/2\), at values of \(\nu_g\) located exactly in between the g-mode eigenvalues \(\left\{\nu_{g,i}\right\}\). For these to be representative of the derivatives evaluated at \(\left\{\nu_{g,i}\right\}\), and thus \(\zeta_p(\nu_i)\), the curvature of these level sets must be heuristically ``small'', in the sense that these step sizes \(h\) must oversample such curvature. In the limit of weak coupling, the avoided crossings between adjacent \(f_n\) become increasingly sharply curved over vanishingly small zones of avoidance (\cref{fig:q}). As such, the length scale on this diagram for such oversampling must be set by the sizes of these zones of avoidance.

\edit1{We find this characteristic scale by expanding the trigonometric terms appearing in \cref{eq:zeta2} in the neighbourhood of $\nu_p$ \cite[following][]{jcd_kepler_2012} to obtain}
\begin{equation}
  \zeta_p(\nu) \sim \left(1 + {\nu^2 \Delta\Pi \over q \Delta\nu}{1 \over 1 + {\pi^2 \over q^2 (\Delta\nu)^2}(\nu - \nu_p)^2}\right)^{-1}; \label{eq:lorentzian}
\end{equation}
\edit1{that is to say,} in the neighbourhood of p-mode eigenvalues \(\nu_p\), the function \({1 \over \zeta_p(\nu_p + \delta\nu)} - 1\) is extremely well approximated by a Lorentzian profile of the form \({A \over \delta\nu^2 + \gamma^2}\), with a characteristic resonance width \(\gamma = q \Dnu / \pi\). If the available mixed-mode period differences are to adequately oversample this width, we therefore require
\begin{equation}
    D_\text{samp} = {\gamma \over h} \sim {2 q \Dnu \over \pi \nu^2 \Delta\Pi_l} = {2 \over \pi} q \mathcal{N}_l \gg 1, \label{eq:sampling}
\end{equation}
\edit1{where $\mathcal{N}_l(\nu) = \Delta\nu / \nu^2 \Delta\Pi_l$ is the relative density of g-modes to p-modes at frequency $\nu$. Correspondingly, when $D_\text{samp} \ll 1,$ we claim that no meaningful constraint on the coupling strength $q$ may be obtained from the mode frequencies alone.}

\begin{figure}[htbp]
  \centering
  \annotate{\includegraphics{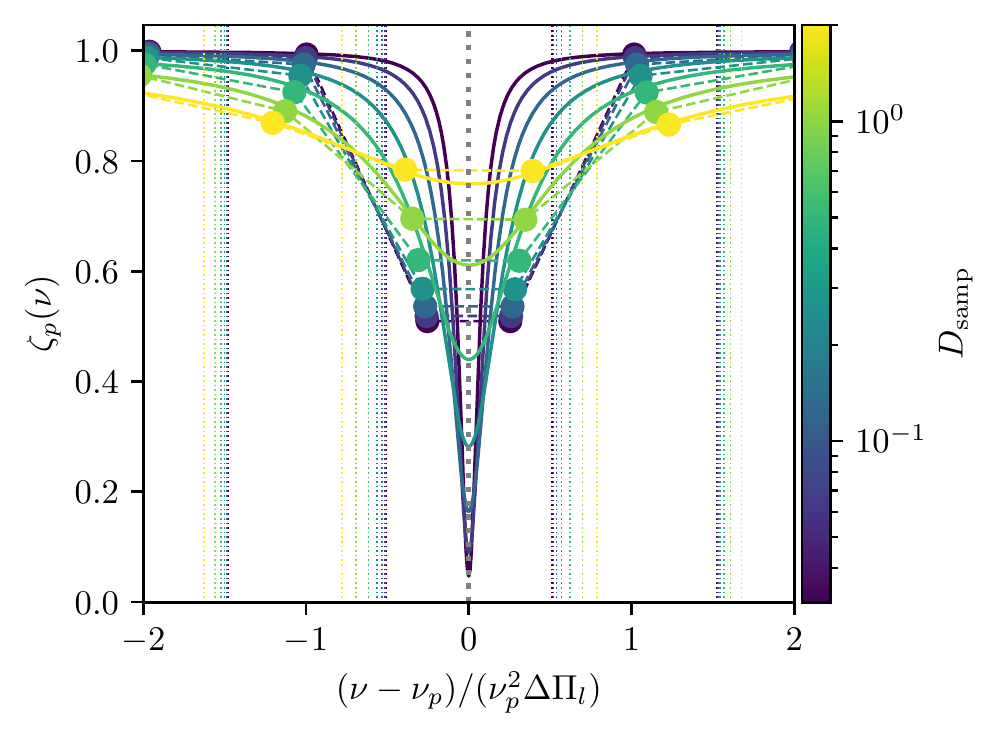}}{\node at (.2, .25){\textbf{(a)}};}
  \annotate{\includegraphics{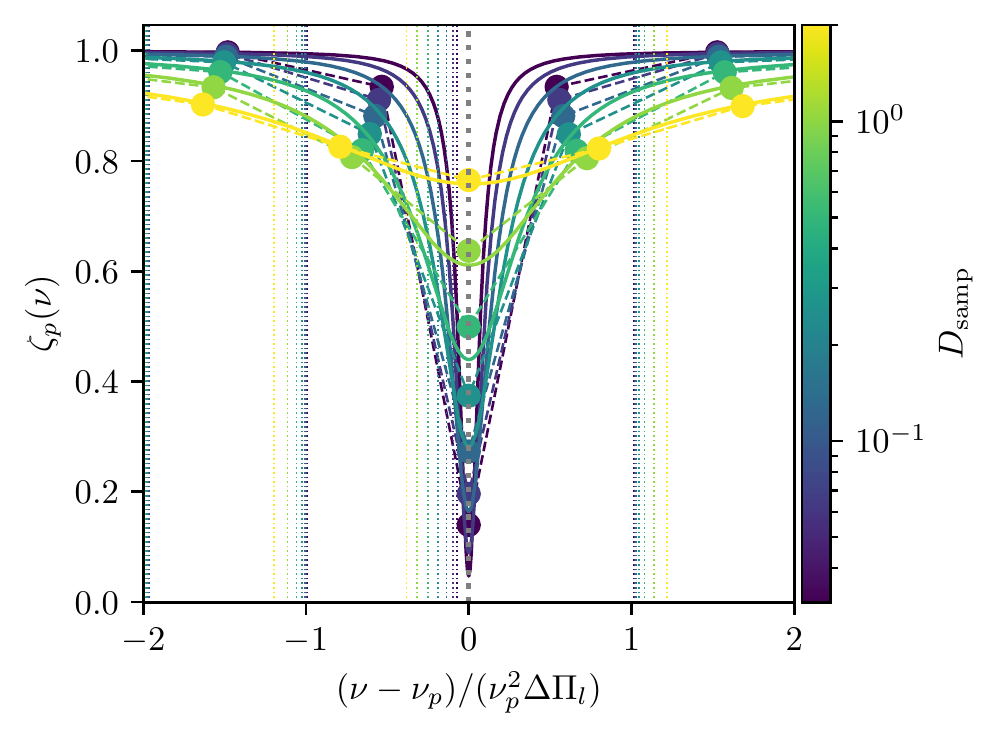}}{\node at (.2, .25){\textbf{(b)}};}
  \caption{Sampling of $\zeta$ by various configurations of mixed modes over a range of $D_\text{samp}$. Solid curves show the function $\zeta_p$ evaluated with values of $q$ associated with the corresponding values of $D_\text{samp}$, indicated with our colour map, in the neighbourhood of a p-mode frequency, $\nu_p$ (indicated with a vertical gray dotted line). One p-dominated mixed mode always exists at that frequency (not shown). Coloured vertical dotted lines show the locations of the remaining g-dominated mixed modes. Finally, filled circles and dashed lines show the scaled mixed-mode period differences, $\delta P / \Delta\Pi_l$, notionally sampling $\zeta_p$ at the midpoints of the mixed-mode frequencies. We plot these quantities for logarithmically-spaced $D_\text{samp} \in \left\{2^{-5}, 2^{-4}, \ldots, 2^{1}\right\}$, \textbf{(a)} for g-modes which are exactly out of resonance, and \textbf{(a)} with one g-mode exactly in resonance, with an underlying p-mode.}
  \label{fig:undersampling}
\end{figure}

\edit1{To understand the basis for this claim, it} is instructive to consider what happens when the sampling criterion \cref{eq:sampling} is violated, which happens in the limit of weak coupling as \(q \to 0\). \edit1{We first consider the case where a p-mode is maximally out of resonance with the available g-modes ($\sin \Theta_g(\nu_p) = 1$), which we illustrate in \cref{fig:undersampling}a. As $D_\text{samp} \to 0$ (and so likewise $q$), the g-dominated mixed mode frequencies tend asymptotically to those of the underlying g-modes, and likewise the p-dominated mixed mode frequencies to the underlying p-modes. We show both the true function $\zeta_p$ generating these mixed modes (coloured solid curves), as well as the values obtained using the finite-difference approximation, \cref{eq:requiredproperty}, with perfect knowledge of $\Delta\Pi_l$ (filled circles joined with dashed lines). The finite-difference approximation is such that these are assumed to sample $\zeta_p$ exactly in between the mixed-mode frequencies.}

\edit1{At $D_\text{samp} \gtrsim 1$, the finite-difference approximation can be seen to hold well. However, as $D_\text{samp} \to 0$, the two period differences closest to the p-mode frequency must tend to $\delta P \to \Delta\Pi_l / 2$, sampled at $\nu_\pm \sim \nu_p \pm \nu_p^2 \Delta\Pi_l / 4$. Thus, were \cref{eq:requiredproperty} taken at face value, one would assume to have sampled $\zeta = {1 \over 2}$ at locations which become increasingly discrepant from the true half-width of the negative peak (which instead also shrinks to $0$ as $q \to 0$). Consequently, one would obtain a systematic overestimate of $q$, which persists across various methods used to estimate it. For example, were \cref{eq:zeta2} or \cref{eq:lorentzian} to be fitted against the period differences, a larger value of $q$ than actually present would be required to reproduce the width of the feature implied by the finite-difference approximation. Alternatively, if the ratio $Z = \zeta_\text{min} / \zeta_\text{max}$ were to be used \cite[\edit2{as in }e.g.][]{farnir_eggmimosa_2021}, we see as $q \to 0$ that one would obtain from finite differences $Z \to {1 \over 2}$, rather than $\to 0$ as required, again leading to a systematic overestimate.}

\edit1{Let us now also examine the near-resonance configuration shown in \cref{fig:axis}. \citet{jcd_kepler_2012} derives several results for this configuration which we will find useful later. Adapting them} to our notation, we have \edit1{that the difference between the two closest mixed modes to resonance can be found as}
\begin{equation}
\begin{aligned}
    \nu_+ - \nu_- &\sim \sqrt{(\nu_p - \nu_{g,0})^2 + 2 \nu_{g,0}(\nu_p + \nu_{g,0}){q \Dnu \Delta\Pi_l \over \pi^2} + \mathcal{O}(q^2)} \\ &\to |\nu_p - \nu_{g,0}|\text{ as } q \to 0. \label{eq:diffasymp}
\end{aligned}
\end{equation}
\edit1{This} separation is bounded from below \edit1{by the case of exact resonance, where} \(\nu_p \to \nu_g \to \nu_c\), as
\begin{equation}
    \nu_+ - \nu_- \ge \delta\nu_\text{res, asy} = {2 \nu_c \over \pi} \sqrt{q \Dnu \Delta\Pi_l}  + \mathcal{O}(q^{3/2}).\label{eq:bound1}
\end{equation}
\edit1{Inserting this into \cref{eq:lorentzian} yields the mixing fraction of both modes,
\begin{equation}
  \zeta_\pm = \zeta_p(\nu_\pm) = \left(1 + {1 \over 1 + q \mathcal{N}_l}\right)^{-1},\label{eq:zetahalf}
\end{equation}
which can be seen to tend to ${1\over2}$ in the limit of weak coupling.}

\edit1{We show in \cref{fig:undersampling}b the same quantities as in \cref{fig:undersampling}a, in this case of exact resonance, where $\nu_{g,0} \to \nu_p$. Again, we can see that while the finite-difference approximation holds well for $D_\text{samp} \gtrsim 1$, its limitations at low $q$ also conspire to give the appearance of stronger coupling than actually present, in much the same fashion as we have described in the out-of-resonance configuration. Nonetheless, these overestimates can be seen not to be as extreme. For example, \cref{eq:lorentzian} and \cref{eq:bound1} together indicate that while $Z$ will still be systematically overestimated from period differences, it does eventually tend to 0 as $q \to 0$.}

\edit1{As such, estimating $\zeta$ from only period differences will yield overestimates of $q$ in configurations of both exact resonance and exact antiresonance with the p-modes, when $D_\text{samp}$ is small. Supposing that intermediate proximity to resonance between the two configurations that we have considered above will yield intermediate overestimates of $q$, we make the stronger claim that when $D_\text{samp} \ngtr 1$, values of $q$ which may be inferred from only mixed-mode frequencies will be systematically higher than required to actually describe their mode coupling. The amount by which $q$ will be overestimated then depends on the extent to which the observationally available p- and g-modes are variously close to or out of resonance\edit2{, as well as the value of $D_\text{samp}$ itself.}}

\hypertarget{algebraic-construction}{%
\subsection{Algebraic Construction}\label{algebraic-construction}}

\edit1{We now examine the near-resonance configuration using the algebraic construction, in which} the rotating mixed-mode frequencies are fully described by a quadratic Hermitian eigenvalue problem
\begin{equation}
    \left(\omega^2
    \begin{bmatrix}
    \mathbb{I}_p & \mathbf{D} \\
    \mathbf{D}^\dagger & \mathbb{I}_g
    \end{bmatrix}
    + 2 \omega \mathbf{R}
   - \begin{bmatrix}
    \mathbf{\Omega}_p^2 & \bm{\alpha} \\ \bm{\alpha}^\dagger & \mathbf{\Omega}_g^2
    \end{bmatrix}
    \right)
    \begin{bmatrix}
    \mathbf{c}_p \\ \mathbf{c}_g
    \end{bmatrix} = 0, \label{eq:rotmatrix}
\end{equation}
as written in block matrix form, where \(\mathbb{I}_p\) and \(\mathbb{I}_g\) are identity matrices of the same rank as the \edit1{notional} pure p- and g-mode subspaces under consideration; \(\mathbf{\Omega}_p\) and \(\mathbf{\Omega}_g\) are diagonal matrices containing the pure p- and g-mode angular frequencies; $\mathbf{R}$ specifies the rotational properties of the star, and \(\bm{\alpha}\) and \(\mathbf{D}\) are matrices describing the pairwise interactions between each combination of p- and g-modes. In \obb{} we show that the rotation matrix $\mathbf{R}$ is well-approximated as being diagonal in the basis of isolated p- and g-modes. This being the case, one may approximate this problem \edit1{in this basis} by solving the simpler Generalised Hermitian Eigenvalue Problem separately for each combination of $m$ and $l$, in the form
\begin{equation}
    \begin{bmatrix}
    \mathbf{\Omega}_p^2 & \bm{\alpha} \\ \bm{\alpha}^\dagger & \mathbf{\Omega}_g^2
    \end{bmatrix}
    \begin{bmatrix}
    \mathbf{c}_p \\ \mathbf{c}_g
    \end{bmatrix}
     = \omega^2
    \begin{bmatrix}
    \mathbb{I}_p & \mathbf{D} \\
    \mathbf{D}^\dagger & \mathbb{I}_g
    \end{bmatrix}
    \begin{bmatrix}
    \mathbf{c}_p \\ \mathbf{c}_g
    \end{bmatrix}, \label{eq:matrix}
\end{equation}
with the diagonal elements of $\mathbf{\Omega}_i$ suitably modified to account for the rotational perturbation associated with the corresponding $l$ and $m$.

\edit1{While this construction is entirely  general, numerical prescriptions} for evaluating these matrix elements have been provided in \citet{ong_semianalytic_2020} as overlap integrals between notional ``\(\pi\)'' and ``\(\gamma\)'' modes \citep[in the sense of][]{aizenman_avoided_1977}. In particular, \edit1{given their prescription for isolating such $\pi$ and $\gamma$ modes,} the coupling matrix element for the \(i\)\textsuperscript{th} p-mode coupling to the \(j\)\textsuperscript{th} g-mode can be written as
\begin{equation}
    \alpha_{ij} = \int \mathrm{d} m\ \bm{\xi}_{\pi,i} \cdot \left(N^2 \mathbf{e}_r \otimes \mathbf{e}_r + \omega_{\pi, i}^2\right) \cdot \bm{\xi}_{\gamma,j},\label{eq:alpha}
\end{equation}
where \(\mathbf{e}_r\) is the unit vector in the radial direction, \edit1{$\otimes$ denotes the tensor product}, \(\bm{\xi}_k\) is the Lagrangian perturbation eigenfunction of the \(k\)\textsuperscript{th} mode, and \(N^2\) is the (squared) buoyancy frequency. Within this construction, the mixed-mode frequencies are found as the eigenvalues of \cref{eq:matrix}. The eigenfunctions of the resulting mixed modes can be written as linear combinations of these \(\pi\) and \(\gamma\) modes:
\begin{equation}
    \bm{\xi} \sim \sum_i c_{p, i} \bm{\xi}_{\pi,i} + \sum_j c_{g,j} \bm{\xi}_{\gamma,j},
\end{equation}
where these coefficients \(c\) are specified by the eigenvectors corresponding to each of these eigenvalues.

For these evolved red giants in particular, two simplifying approximations are typically made in this analysis. The first is \edit1{to reduce \cref{eq:matrix} to a conventional Hermitian eigenvalue problem. This may be done \citep[following][]{lavely_effect_1992} by observing that all of the \(|D_{ij}| \ll 1\), and thus expanding the equivalent Hermitian eigenvalue problem in powers of $D$ to yield an effective coupling strength}
\begin{equation}
  A_{ij} \sim \alpha_{ij} - {1\over 2}\left(\omega_i^2 + \omega_j^2\right) D_{ij} + \mathcal{O}(D^2).
\end{equation}
\edit1{We truncate this expansion to zeroth order in our analytic development here \citep[as in e.g.][]{deheuvels_insights_2010} for simplicity of notation, but will use the first-order correction in our later numerical calculations.}

The second \edit1{approximation} is that the coupling strengths are sufficiently weak that each g-dominated mixed mode may be assumed to result from the two-mode coupling between a single g-mode and a single p-mode. Following \citet{ong_semianalytic_2020}, \cref{eq:matrix} may be written in that case as
\begin{equation}
    \begin{bmatrix}
    \omega_p^2 & \alpha \\ \alpha & \omega_g^2
    \end{bmatrix}
    \begin{bmatrix}
    1 \\ u_\pm
    \end{bmatrix}
     \sim \omega_\pm^2
    \begin{bmatrix}
    1 & 0 \\
    0 & 1
    \end{bmatrix}
    \begin{bmatrix}
    1 \\ u_\pm
    \end{bmatrix}, \label{eq:matrix2}
\end{equation}
where the frequency eigenvalues are found as
\begin{equation}
    \omega_\pm^2 \sim {\omega_p^2 + \omega_g^2 \over 2} \pm {1\over2}\sqrt{\left(\omega_p^2 - \omega_g^2\right)^2 + 4 \alpha^2},
\end{equation}
with corresponding eigenvectors specified by
\begin{equation}
    u_\pm = {\omega_\pm^2 - \omega_p^2 \over \alpha} = {\alpha \over \omega_\pm^2 - \omega_g^2}. \label{eq:u}
\end{equation}
These eigenvalue components can be used to define mixing fractions \(\zeta\) as
\begin{equation}
    \zeta_\pm \equiv {u_\pm^2 \over 1 + u_\pm^2} = {|c_\gamma|^2 \over |c_\pi|^2 + |c_\gamma|^2}. \label{eq:matzeta}
\end{equation}
\citet{ong_semianalytic_2020} demonstrate that this is equivalent to the original definition of mixing fractions as inertia ratios, in that it reduces to \cref{eq:ratios} in the JWKB approximation. However, this construction is also generally applicable to mixed modes outside of the JWKB limit.

Of these two solutions, one of them will be closer in frequency to the g-mode than the other; it will have frequency \(\omega_+\) if \(\omega_g > \omega_p\), and vice versa. Far from resonance, this frequency, which we denote by \(\omega\), again can be thought of as a function of the associated g-mode frequency \(\omega_g\) (\cref{fig:axis}). We may verify, as above, that
\begin{equation}
    \zeta \sim {\partial \omega \over \partial \omega_g} \label{eq:matrixspacing}
\end{equation}
\edit2{\citep[cf. Appendix B of][and \autoref{app:verify}]{ong_semianalytic_2020}}.
Since this property is also shared by the continuous construction resulting from asymptotic analysis, we conclude, conversely, that such an asymptotic construction operates best in the same regime as where three- and higher-mode coupling may be ignored algebraically. We formalise this statement more rigorously in \autoref{sec:reduction}.

Again, we evaluate properties of the resonant pair. We note that
\begin{equation}
    \alpha \left(u_{\pm} + {1 \over u_{\pm}}\right) = \pm{\alpha \over \sqrt{\zeta_\pm(1-\zeta_\pm)}} = 2 \omega_{\pm}^2 - \omega_p^2 - \omega_g^2 = \pm \left(\omega_+^2 - \omega_-^2\right).
\end{equation}
Moreover,
\begin{equation}
    \omega_+^2 - \omega_-^2 = 2 \sqrt{\left(\omega_p^2 - \omega_g^2 \over 2\right)^2 + \alpha^2}. \label{eq:algdiff}
\end{equation}
Eliminating the coupling strength \(\alpha\) \edit1{in favour of \(\zeta\)} thus yields
\begin{equation}
    \omega^2_+ - \omega^2_- = {|\omega_p^2 - \omega_g^2| \over \sqrt{1 - 4 \zeta(1 - \zeta)}}.
\end{equation}
Note that it does not matter whether \(\zeta_+\) or \(\zeta_-\) (associated with the lower or higher root) are used in this expression, since \(\zeta_+ = 1 - \zeta_-\) by orthogonality. Thus,
\begin{equation}
    \nu_+ - \nu_- = {|\nu_p - \nu_g| \over \sqrt{1 - 4\zeta(1-\zeta)}} \ne \nu^2 \zeta \Delta\Pi_l,\label{eq:violation}
\end{equation}
again in violation of \cref{eq:requiredproperty}. Finally, we once again consider the case where \(\nu_p\) and \(\nu_g\) are exactly in resonance at frequency \(\nu_c\). In that case, \(\zeta = 1/2\) for both modes \edit1{(compare \cref{eq:zetahalf}), and \cref{eq:violation} becomes singular. Rather, we obtain from \cref{eq:algdiff} that}
\begin{equation}
    \nu_+ - \nu_- \ge \delta\nu_{\text{res}, \text{alg}} = {\alpha \over 4\pi^2 \nu_c}. \label{eq:bound2}
\end{equation}

\subsection{A Weak-Coupling Limit for the Nonasymptotic Mixed-mode Eigenvalue Equation}
\label{sec:reduction}

While both the nonasymptotic algebraic and asymptotic constructions appear to yield good descriptions of the phenomenology of the observed mixed modes, little discussion has taken place regarding how the two are related. \edit1{A better understanding of how the two formulations are related would both significantly allieviate some known computational and conceptual difficulties, as well as ease the interpretation of existing observational results in comparison to numerical models of stellar structure. In particular, the matrix construction becomes increasingly unwieldy to apply for evolved red giants, where the relevant coupling matrices become very large \citep{ong_surface_1}; numerical solutions to the matrix eigenvalue problem, which scale in time complexity with the rank $N$ of these matrices as $\mathcal{O}(N^3)$, become prohibitively expensive. At the same time, despite extensive observational characterisation, the asymptotic coupling strength is difficult to relate directly to mode frequencies obtained from numerical solutions to the pulsation problem from stellar modelling. Should some equivalence between the two formulations be found, it would allow measurements of the quantities of one formulation to be interpreted and used within the other.}

The quantity $q$ carries an interpretation of a product of transmission coefficients in the JWKB limit of the matrix construction \cite[\S 3.3]{ong_semianalytic_2020}, \edit1{but this is not easy to relate to other quantities in the algebraic construction, which} primarily depends on global, rather than local, properties of the wavefunctions. \edit1{However, we note that the mixing fractions and the mode frequencies of both constructions tend to each other in the limit $q \to 0$. If we further demand that their avoided crossings have the same shape at both small $q$ and small $\alpha$, then we may also identify the minimum separations of the avoided crossings, \cref{eq:bound1,eq:bound2}, that appear in both constructions when the underlying p- and g-modes are exactly in resonance. Doing so then allows us to effect a translation between the coupling strengths of the algebraic and asymptotic constructions:}
\begin{equation}
    q \sim {1 \over \Dnu\ \Delta\Pi_l}\left(\alpha \over 8 \pi \nu^2\right)^2.\label{eq:translate}
\end{equation}
\edit1{The oversampling criterion, \cref{eq:sampling}, may therefore be rewritten in terms of \(\alpha\) as}
\begin{equation}
    D_\text{samp} = {2 \over \pi}\left({\alpha \over 8 \pi \nu^3 \Delta\Pi_l}\right)^2 \gg 1. \label{eq:sampling2}
\end{equation}

\edit1{Building on this result, we} will now go further and derive a weak-coupling limit for the \edit1{eigenvalue equation of the} algebraic construction, and show that it passes naturally to \edit1{that of} the asymptotic construction. At the heart of this discussion is the asymptotic eigenvalue equation, \cref{eq:eig}, which we will rewrite as
\begin{equation}
  F_\text{asy}(\nu) = \sin\Theta_p(\nu)\sin\Theta_g(\nu) - q \cos\Theta_p(\nu)\cos\Theta_g(\nu)\label{eq:eig-trig}
\end{equation}
so as to avoid having any singular points, while retaining the same roots. The nonasymptotic analogue to this is the characteristic equation of the Hermitian eigenvalue problem,
\begin{equation}
  f(\omega) = \det
  \begin{bmatrix}
  \omega^2 \mathbb{I} - \mathbf{\Omega}_p^2 & \omega^2 \mathbf{D} - \bm{\alpha} \\ 
  \omega^2 \mathbf{D}^\dagger - \bm{\alpha}^\dagger & \omega^2 \mathbb{I} - \mathbf{\Omega}_g^2
  \end{bmatrix} \equiv \det \mathbf{L} = 0. \label{eq:mat-orig}
\end{equation}
Without loss of generality, we may obtain an equivalent expression with the same roots by taking the determinant of some other matrix $\mathbf{P L Q}$ of equal rank, so long as $\det \mathbf{P}(\omega)$ and $\det \mathbf{Q}(\omega)$ have no roots for $\omega > 0$. This being the case, we choose instead as our characteristic equation that
\begin{equation}
  F_\text{mat}(\omega) = \det
  \begin{bmatrix}
  - \mathbb{I} + \omega^2\mathbf{\Omega}_p^{-2} & \mathbf{\Omega}_p^{-1}(\omega \mathbf{D} - {1 \over \omega}\bm{\alpha}) \\ 
  (\omega \mathbf{D}^\dagger - {1 \over \omega}\bm{\alpha}^\dagger)\mathbf{\Omega}_p^{-1} & \mathbb{I} - {1 \over \omega^2}\mathbf{\Omega}_g^2
  \end{bmatrix} = 0.\label{eq:mat-char}
\end{equation}
Let us first restrict our attention to the diagonal blocks of this matrix, which are both also diagonal matrices. The determinants of these blocks are thus products of the diagonal entries: the determinant of the upper left block takes the form
\begin{equation}
\det \left(\mathbb{I} - \omega^2 \mathbf{\Omega}_p^{-2}\right) = \prod_{n_p}\left(1 - {\omega^2 \over \omega^2_{p, n_p}}\right) \sim \prod_{n_p}\left(1 - \left(\nu / \Delta\nu \over n_p + {l \over 2} + \epsilon_{p,nlm}\right)^2\right),\label{eq:detp}
\end{equation}
while that of the lower right block is of the form
\begin{equation}
\det \left(\mathbb{I} - \mathbf{\Omega}_g^2 \cdot {1 \over \omega^2}\right) = \prod_{n_g}\left(1 - {\omega_{g, n}^2 \over \omega^2}\right) \sim \prod_{n_g}\left(1 - \left(1 / \nu\Delta\Pi_l \over n_g + \epsilon_{g,nlm}\right)^2\right),\label{eq:detg}
\end{equation}
where we have used the asymptotic relations, \cref{eq:p,eq:g}.

We now recall that analytic functions admit Weierstrass factorisations as infinite products over their roots. In particular, the trigonometric functions may be factorised as
\begin{equation}
  \sin \pi x  = \pi x \prod_{n=1}^{\infty} \left(1 - \left(x\over n\right)^2\right);~~ \cos \pi x  = \prod_{n=0}^{\infty} \left(1 - \left(x\over n + {1 \over 2}\right)^2\right),
\end{equation}
which is clearly very similar in structure to \cref{eq:detp,eq:detg}. Motivated by this, we define in trigonometric form the characteristic functions
\begin{equation}
\begin{aligned}
  F_p(\omega) &\equiv \det \left(\mathbb{I} - \omega^2 \mathbf{\Omega}_p^{-2}\right) \sim \sin \Theta_p(\omega)\\
  F_g(\omega) &\equiv \det \left(\mathbb{I} - \mathbf{\Omega}_g^2 \cdot {1 \over \omega^2}\right) \sim \sin \Theta_g(\omega)\label{eq:pg-trig}
\end{aligned}
\end{equation}
for the p- and g-mode cavities separately. Note that such trigonometric definitions are in principle correct only up to multiplication by some functions, $f_p(\omega)$ and $f_g(\omega)$\edit1{, having no zeros for $\omega > 0$}. However, we may always choose a different transformation to \cref{eq:mat-orig} (rather than \cref{eq:mat-char}), scaling each block by some \edit1{appropriate} function of frequency, without changing the roots of these characteristic functions. In particular we may choose such a transformation that that sets these functions to 1. In turn, our subsequent expressions \edit1{turn out not to} depend on the precise transformation chosen. We therefore may assume $f_p$ and $f_g$ both to be 1 without loss of correctness.

Let us now consider \cref{eq:mat-char} again, to which we apply the identity
\begin{equation}
  \det \begin{bmatrix}
  \mathbf{A} & \mathbf{B} \\ \mathbf{C} & \mathbf{D}
  \end{bmatrix} = \det \mathbf{A}\ \det \mathbf{D}\ \det \left(\mathbb{I} - \mathbf{D}^{-1}\mathbf{C}\mathbf{A}^{-1}\mathbf{B} \right),
\end{equation}
and expand the final multiplicand in powers of the coupling matrices as
\begin{equation}
   \det \left(\mathbb{I} + \lambda \mathbf{A}\right) = 1 + \lambda \mathrm{Tr}\ \mathbf{A} + \mathcal{O}(\lambda^2).
 \end{equation}
In this fashion we obtain
\begin{equation}
\begin{aligned}
  F_\text{mat}(\omega) &\sim F_p(\omega)F_g(\omega)\left(1 - \mathrm{Tr} \left[(\omega^2 \mathbb{I} - \mathbf{\Omega}^2_g)^{-1} \mathbf{A}^\dagger (\omega^2 \mathbb{I} - \mathbf{\Omega}^2_p)^{-1} \mathbf{A}\right] + \ldots\right)\\
  &= F_p(\omega)F_g(\omega) - \sum_{i, j}{F_p(\omega)F_g(\omega)A_{ij}A_{ji} \over (\omega^2 - \omega_{p, i}^2)(\omega^2 - \omega_{g, j}^2)}+ \ldots
\end{aligned} \label{eq:eigsum}
\end{equation}
where for compactness we write $\alpha_{ij} - \omega^2 D_{ij} \equiv A_{ij}$. To simplify the sum, we consider the derivatives of \cref{eq:detp,eq:detg} with respect to $\omega$. By the product rule, this yields
\begin{equation}
  {\partial \over \partial \omega} F_p(\omega) = \sum_{n_p} {2 \omega F_p(\omega) \over \omega^2 - \omega_{n_p}^2};~~ {\partial \over \partial (-1/\omega)} F_g(\omega) = \sum_{n_p} {(2 \omega \omega_{n_g}^2) F_p(\omega) \over \omega^2 - \omega_{n_g}^2}.
\end{equation}
Accordingly, we rewrite the sum in \cref{eq:eigsum} in the fairly suggestive form
\begin{equation}
\sum_{ij} {2 \omega F_p(\omega) \over \omega^2 - \omega_{p,i}^2} \cdot {2 \omega \omega_{g,j}^2 F_g(\omega) \over \omega^2 - \omega_{g, j}^2} \cdot {|A_{ij}|^2 \over 4 \omega^2 \omega_{g,j}^2}.
\end{equation}
So far we have made no assumptions about the structure of the matrices $\bm{\alpha}$ and $\mathbf{D}$. At this point, however, let suppose that the quantity
\begin{equation}
  Q_{ij}(\omega) = {\pi^2 \over \Delta\nu\ \Delta\Pi_l}\left(|A_{ij}| \over 2 \omega \omega_{g,j}\right)^2,
\end{equation}
which we have defined by adapting \cref{eq:translate}, may be treated as being numerically independent of its matrix indices. In that case it can be brought out of the sum, which collapses to yield
\begin{equation}
\begin{aligned}
  F_\text{mat} &= F_p(\omega) F_g (\omega) - Q\left({\Delta\nu \over \pi}{\partial \over \partial \nu} F_p\right) \left({\Delta\Pi_l \over \pi}{\partial \over \partial (-1/\nu)} F_g\right) + \mathcal{O}(Q^2)\\
  &= F_p(\omega) F_g (\omega) - Q\left({\partial \over \partial \Theta_p} F_p\right) \left({\partial \over \partial \Theta_g} F_g\right) + \mathcal{O}(Q^2). \label{eq:equiv}
\end{aligned}
\end{equation}
We can now see that identifying $Q$ with $q$, and substituting the trigonometric expressions \cref{eq:pg-trig} into \cref{eq:equiv}, does indeed cause the algebraic construction to yield the asymptotic eigenvalue equation, \cref{eq:eig-trig}. Moreover, our explicit construction allows us to make more precise statements regarding the conditions under which the two formulations become equivalent. In particular, we require that (1) the elements of the coupling matrices have no parametric dependence on the p-mode frequencies, and depend on the g-mode frequencies as $(\alpha_{ij} - \omega^2 D_{ij}) \propto \omega_{g,j}$; and that (2) we may neglect terms to fourth and higher order in the coupling matrices (i.e. the coupling between mode cavities is weak). In general, however, it is known that condition (1) is not satisfied by stellar models: the coupling matrix elements can be shown at least numerically to exhibit dependence on both the p- and g-mode indices, and their dependence on the g-mode frequency is not necessarily of the required form (as we will examine below). \edit1{In a local sense, an effective value of $q$ may be obtained using \cref{eq:translate} from the on-resonance matrix element corresponding to each avoided crossing considered in isolation, yielding frequency dependence of the kind considered in e.g. \citet{farnir_eggmimosa_2021,jiang_evolution_2022}. However,} absent some conspiracy of stellar structure, the two formulations are globally inequivalent.

\hypertarget{numerical-results}{%
\section{Numerical Results}\label{numerical-results}}

We have now \edit1{examined how the finite-difference approximation} \cref{eq:requiredproperty} breaks down when the avoided crossings are not adequately oversampled. \edit1{We quantify this using an oversampling discriminant, expressed with respect to the coupling strength in both the asymptotic (\cref{eq:sampling}) and algebraic (\cref{eq:sampling2}) formulations. When this oversampling requirement is not satisfied, the coupling strengths $q$ may be systematically overestimated.} The mixing fraction \(\zeta\) --- inferred values of which depend on $q$ --- is itself an observationally significant diagnostic quantity in its own right, and errors in estimating it will in turn contaminate the interpretation of other seismic measurements, such as rotation rates (e.g.~via \cref{eq:rotation}). As such, it is important to understand the extent to which existing estimates of it, and by extension quantities derived from it, remain reliable in the face of these systematic issues.

\subsection{Stellar Modelling}

We first assess this qualitatively, by examining how well the oversampling criterion is satisfied numerically by stellar models produced with evolutionary calculations. For the purposes of our following discussion, we will use the relative density of g-modes to p-modes at \(\numax\), \(\mathcal{N}_l(\numax) = \Dnu / \numax^2 \Delta\Pi_l\), as an indicator of evolution up the red giant branch. In particular, we focus on the region \(\mathcal{N}_{l = 1}(\numax) < 30\), which encompasses essentially the entirety of the current observational sample of \edit1{first-ascent} red giants for which measurements of \(\zeta\) using asymptotic techniques have been attempted \citep[e.g.][]{mosser_period_2018, gehan_core_2018}.

\edit1{Since we seek to assess whether or not the oversampling criterion \cref{eq:sampling} holds at all, it would be circular to evaluate it with respect to some phenomenological $q$ as inferred from model mixed modes, as we have earlier established that we may only meaningfully do so if this criterion were to be already satisfied in the first place. Moreover, the asymptotic construction in which $q$ appears is itself} strictly speaking not universally applicable to mixed-mode oscillators in general, as the eigenvalue equation from which it is derived, \cref{eq:eigjwkb}, further relies on the JWKB approximation, which serves as a good description only at high wavenumber (in this case, at high radial order). The JWKB approximation is known not to hold for both subgiants and early red giants at low $\mathcal{N}_l$ \cite[where the underlying g-modes are of low degree: cf.][]{bedding_replicated_2012}, or for very evolved red giants at high $\mathcal{N}_l$ \cite[where the underlying p-modes are of low degree: cf.][]{stello_nonradial_2014}. Since well-defined bounds for $\mathcal{N}_l$ in which the JWKB approximation may be applied are not generally available, we cannot necessarily guarantee that it holds good within our evolutionary regime of interest. Obversely, our assessment of whether or not the oversampling criterion holds should preferably be independent of the validity of the JWKB approximation per se. \edit1{For these reasons, we will instead evaluate the oversampling discriminant using the nonasymptotic expression, \cref{eq:sampling2}.}

We \edit1{evaluate our quantities with respect to} stellar models in a set of evolutionary tracks produced by the stellar evolution code \mesa{} r15140 \citep{mesa_paper_1, mesa_paper_2, mesa_paper_3, mesa_paper_4, mesa_paper_5}, constructed in the same fashion as those considered in \obb. For each model under consideration, we compute the off-diagonal matrix element  corresponding to the resonant pair of dipole \(\pi\)- and \(\gamma\)-modes closest in frequency to \(\numax\). We do so using the pulsation code \gyre{} \citep{townsend_gyre_2013} and \cref{eq:alpha}.

\begin{figure}[htbp]
    \centering
    \includegraphics[width=.495\textwidth]{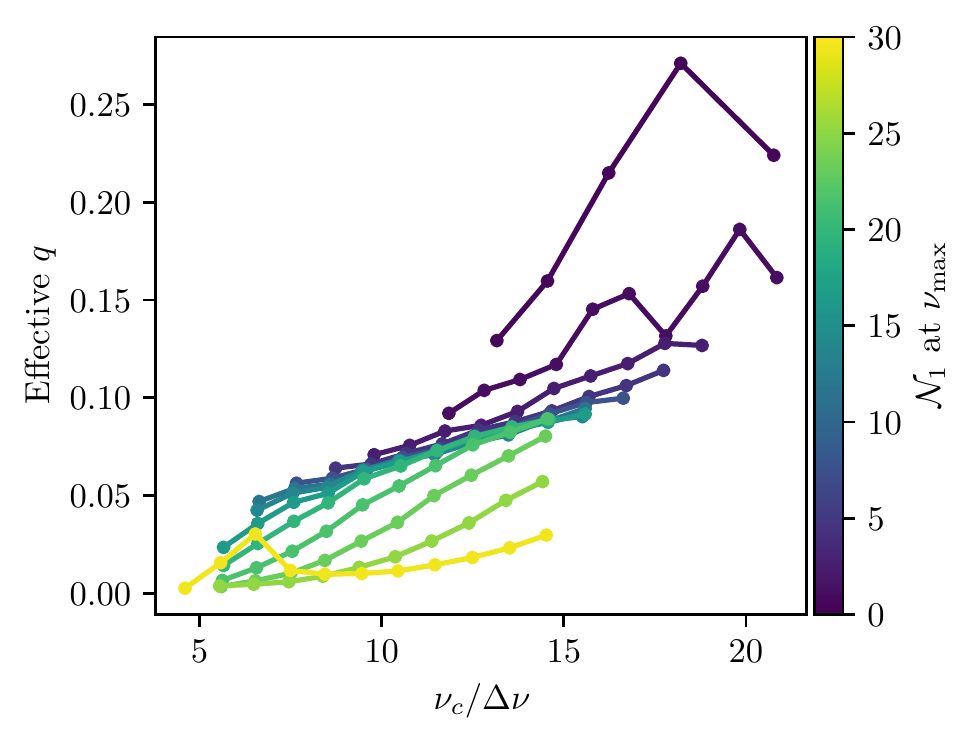}
    \caption{Dependence of the effective coupling strength $q$ as a function of frequency. Values from each stellar model are coloured by $\mathcal{N}_1$, the density of g-modes per p-mode (a proxy for RGB evolution), and joined with lines of the same colour.}
    \label{fig:coupling}
\end{figure}

\edit1{Conventional methods by which $q$ is inferred from mixed modes necessitate} either specifying some parametric form of $q$ as a function of frequency, or, more commonly, ignoring its frequency dependence altogether. \edit1{By evaluating it directly from stellar structure using \cref{eq:translate} instead}, we are in a position to examine \edit1{this frequency dependence without making any such constraining assumptions}. We show these effective coupling strengths in \cref{fig:coupling} as a function of the normalised frequency $\nu / \Delta\nu$, along an evolutionary track of solar composition and $M = 1.4 M_\odot$, for modes within $\pm 5 \Delta\nu$ of \numax. These values are evaluated as a function of $\nu_c$ (i.e. the central frequencies of pairs of g- and p-modes closest to resonance); such resonant-pair frequencies are located close to the p-mode frequencies for evolved red giants, and close to the g-mode frequencies for subgiants. Values for each stellar model are coloured by $\mathcal{N}_1$, evaluated at $\numax$ for each stellar model.

Unlike the constant, linear, or quadratic dependences on frequency that are typically assumed \cite[e.g.][]{cunha_analytical_2019,dreau_dipolar_2020}, the actual variation of the coupling strength with frequency appears quite complicated, and also exhibits further nontrivial morphological variations depending on evolutionary stage. Specifically, we see that it exhibits significant, \edit1{and seemingly oscillatory,} curvature in our subgiant models \citep[which \edit1{will require more complicated analysis than previously considered}, e.g.][\edit1{to explain}; although the \edit1{applicability of the JWKB approximation} to subgiants is \edit1{perhaps} questionable]{pincon_probing_2020}. For more evolved red giants it clearly also varies significantly, increasing by more than a factor of 2 between the lowest and highest p-modes within our frequency range of interest --- this is consistent with the phenomenological findings of \cite{farnir_eggmimosa_2021}\edit1{, performed on a mode-by-mode basis}. At the same time, the functional dependence on frequency appears not to adhere well to the power-law parameterisation suggested by the JWKB analysis of \cite{pincon_probing_2020}\edit1{, which would be incapable of producing inflection points as exhibited by the curve at $\mathcal{N} \sim 20$.}

\begin{figure}
\centering
\annotate{\includegraphics[width=.495\textwidth]{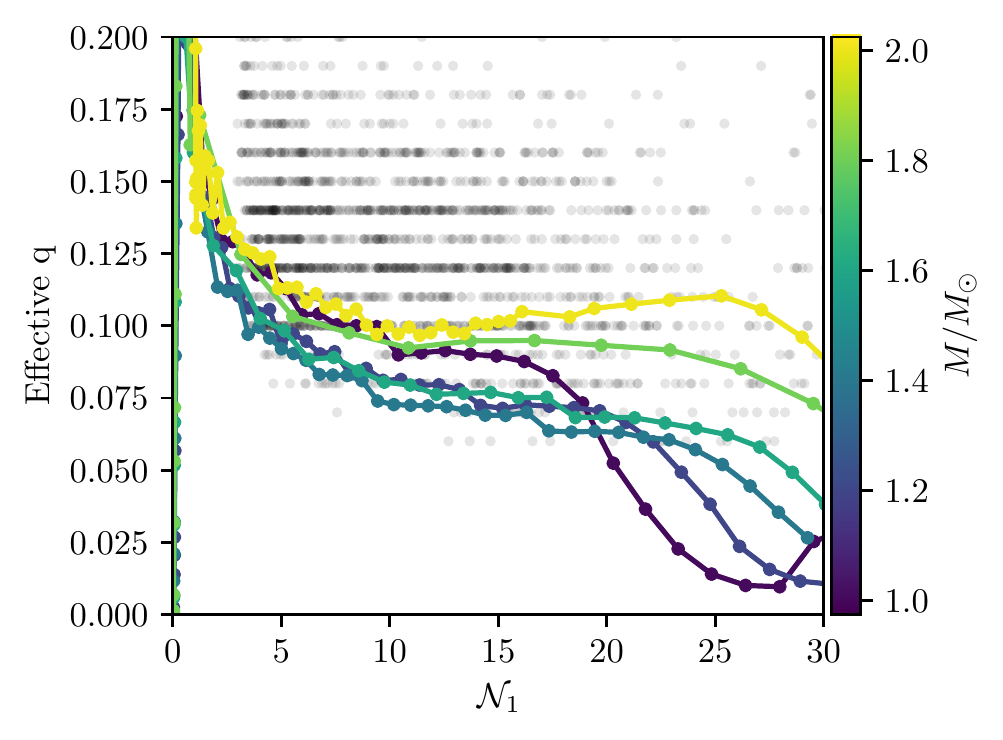}}{\node at (.21, .23){\textbf{(a)}};}
\annotate{\includegraphics[width=.495\textwidth]{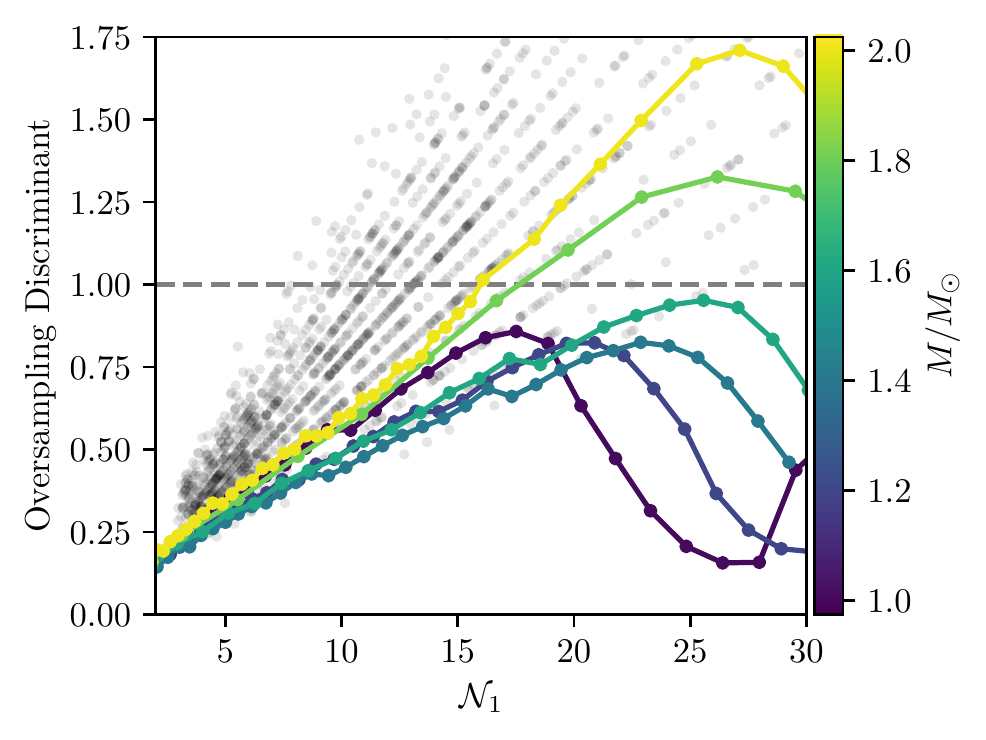}}{\node at (.21, .9){\textbf{(b)}};}
\caption{\textbf{(a)} Effective coupling strength $q$, and \textbf{(b)} sampling discriminant $D_\text{samp}$ (\cref{eq:sampling,eq:sampling2}), evaluated at \(\numax\) for dipole modes (\(l=1\)) associated with tracks of evolutionary models with different initial masses (shown with different colours)\edit1{, and for the first-ascend red giant observational sample of \cite{mosser_period_2017} (shown with gray points)}. These are plotted as a function of \(\mathcal{N}_l = \Dnu / \numax^2\Delta\Pi_l\), the relative density of g-modes to p-modes at \(\numax\), for dipole modes ($l = 1$). In panel b, the critical value of 1 has been marked out with the horizontal dashed line.\label{fig:disc}}
\end{figure}

We show in \cref{fig:disc}a the value of $q(\numax)$ as a representative value for each stellar model, over the course of stellar evolution, for a series of evolutionary tracks ranging from $1M_\odot$ to $2 M_\odot$. We then use it to evaluate the sampling discriminant, \cref{eq:sampling2}, and show the values of this discriminant in \cref{fig:disc}b, following the evolution of these quantities up to $\mathcal{N}_1(\numax) \sim 30$. We find that the various evolutionary tracks exhibit fairly similar morphology in the evolution of this coupling strength. After exhibiting relatively strong mode coupling as subgiants with primarily p-dominated mixed modes, the effective coupling strengths of our model stars level off with increasing $\mathcal{N}_1 \lesssim 15$, before again falling off precipitously. Correspondingly, the sampling discriminant appears to grow linearly with $\mathcal{N}$ until this turnoff point, after which it decreases sharply. Within the range of \(\mathcal{N}_l(\numax)\) and stellar masses that we have considered, the sampling discriminant never takes values much larger than unity for these tracks. On the contrary, it is significantly smaller than unity for most of our \mesa{} models within this range: the oversampling criteria, \cref{eq:sampling,eq:sampling2}, are generally speaking not satisfied by mixed modes near \numax, except for higher masses at values of \(N_1(\numax)\) near the upper end of the observable range. We note that this discriminant does exceed unity near the turnoff point in the evolution of the coupling strengths for stars massive enough to host non-degenerate helium cores, \edit1{and this transition from core electron degeneracy is known to induce other seismic signatures \citep[e.g][]{deheuvels_degeneracy_2022}. However,} it is difficult to say for certain from this cursory analysis whether this \edit1{feature} is of genuine astrophysical origin or merely an interesting numerical coincidence in our stellar modelling\edit1{, and a detailed investigation of this lies beyond the scope of this work.}

\edit1{Since the oversampling criterion $D_\text{samp} > 1$ is not satisfied by our numerical models, we accordingly} conclude that the actual combinations of frequency spacings, period spacings, and coupling strengths associated with first-ascent red giants in this region of parameter space are quite likely insufficiently informative to unambiguously constrain the mixing fractions \(\zeta\) of near-resonance mixed modes without further input, e.g.~from explicit stellar modelling, or strong priors \edit1{on $q$. Following our earlier discussion in \autoref{sec:overestimate}, attempts to infer coupling strengths under these conditions would yield values that are systematic overestimates. To assess the potential impact of this, we also show in both panels of \cref{fig:disc}, with gray circles, the values of $q$ (\ref{fig:disc}a), and the implied oversampling discriminant (\ref{fig:disc}b), of first-ascent red giants from the catalogue of coupling strengths published in \cite{mosser_period_2017}. We see that the values of $q$ returned from our numerical \mesa{} models form a lower envelope to most of the reported data points. Even despite being higher than our numerical predictions, the majority of the corresponding values of $D_\text{samp}$ can still be seen to be less than 1. These features are entirely consistent with the interpretation supplied in \autoref{sec:overestimate}, of observationally inferred values of $q$ being systematically overestimated when $D_\text{samp} < 1$. Supposing that such an interpretation were to be correct, it would imply that the true coupling strengths of these mixed modes would largely be lower than reported, resulting in true values of the oversampling discriminant which would be also lower than those shown here.}

\hypertarget{hare-and-hounds-exercise}{%
\subsection{Hare-and-Hounds Exercise}\label{hare-and-hounds-exercise}}

The primary applications of the stretched-\'echelle-diagram construction have been to make rotational measurements of first-ascent red giants in the regime where the linear expression, \cref{eq:rotation}, does not hold well. \edit1{However, measurements made using these diagrams implicitly assume that the values of $q$ used to make them are consistent with those actually generating the mixed modes being stretched. Our prior discussion suggests that this may not be the case for a large fraction of our observational sample.} Thus, we now seek to quantify the performance of these existing methods for rotational characterisation that use the stretched-\'echelle construction, in the face of \edit1{these systematic issues. In keeping with the overall direction of this series of papers, we also examine prospects for adapting such measurements for recovering quantities required for rotational inversions}.

We do this by way of a hare-and-hounds exercise. For our hares, we use rotating \textsc{mesa} evolutionary models, with angular momentum transport following the prescription of \cite{fuller_slowing_2019}, and the same initial rotational configuration at ZAMS as the evolutionary tracks used in \obb, generated at initial masses of 1.2 through 1.8 $M_\odot$ at a spacing of $0.2M_\odot$. From these models we compute the dipole mixed-mode frequencies, first without isolation into the basis of $\pi$ and $\gamma$ modes, at all azimuthal orders. Additionally, we also then evaluate the associated $\pi$ and $\gamma$-mode eigensystems, thereby obtaining a separate set of linearised rotational-splitting estimates. For each mixed mode, $1-\zeta$ is evaluated by taking the inverse of the ratio of the mode inertia of the mixed mode under consideration, and the value obtained by interpolating the mode inertiae of the $\ell = 1$ $\pi$-modes to the specified mixed-mode frequency:
\begin{equation}
   \zeta_{nlm} = 1 - {I_\pi(\nu_{nlm}) \over I_{nlm}}.
\end{equation}
We select models with $\mathcal{N}_1$ taking values between 2 and 30 from each of these evolutionary track as our primary set of hares, intended to investigate mass dependences of these methodological systematics; we will refer to these as Set I. In addition, we also generated one evolutionary track at $1.4 M_\odot$ with the ZAMS rotation rate being half that of the same track in Set I, and another with double the ZAMS rotation rate, as an additional pair of sets of hares, from which we selected models and generated frequencies and rotation estimators in the same fashion. We will refer to these hares as Set II.

As our hound, we use the template-matching rotational measurement techniques described in \cite{gehan_core_2018}, which are built upon the stretched-\'echelle-diagram construction. These techniques require estimates of the mixed-mode spectrum (in particular, the g-mode period spacing and phase offset, as well as the coupling strength between the mode cavities) to be specified a priori. Rather than simulate an entire fitting and peakbagging pipeline for these quantities, we mimic \edit1{both internal systematic and statistical} errors on them by sampling posterior distributions of $\Delta\Pi_1$, $\epsilon_g$ and constant $q$. We treat as (simulated) observational data the computed mixed $m = 0$ modes (with a nominal frequency measurement uncertainty of 0.005 $\mu$Hz), and evaluate likelihood functions for a generative model for the mixed-mode frequencies (specified via \cref{eq:eig}) that accepts these three quantities as input parameters, in conjunction with the $\pi$-mode frequencies at $m = 0$ (held fixed with no noise for this procedure). Samples from the posterior distributions with uniform sampling were then constructed using \texttt{emcee}. This procedure was used to supply the hound with $q$, $\Delta\Pi_1$, and $\epsilon_g$, drawn from the joint posterior distribution, as well as nominal measurement errors, estimated from the $1\sigma$ quantiles of the marginal posterior distributions. This was done independently for each hare. The hound was also supplied with the rotating mixed-mode frequencies, with simulated measurement errors as above, as well as the nonrotating $\pi$-mode frequencies, whose simulated measurement errors were also inflated by the envelope rotation rates of the stellar models, which are ordinarily ignored in this procedure.

We recall that in the first-order approximation, mode frequencies in a rotating star relate to those in a non-rotating configuration through \cref{eq:rotation}.
As such, relating the observed quantities to structural stellar models on a multiplet-by-multiplet basis\edit1{, as necessary for e.g. rotational inversions,} will require accurate determination of $m$, $\zeta$, and $\delta\omega_\text{rot}$. While current methods adapted for red giants \cite[including that of][]{gehan_core_2018} are able to estimate the first two quantities, they derive only a single aggregate value of $\Omega_\text{core}$ from template-matching against stretched \'echelle diagrams. Nonetheless, we will now assess the robustness of these methods in recovering these quantities when confronted with our hares.

\hypertarget{hare-and-hounds-zeta}{%
\subsubsection{Inference of $\zeta$}\label{hare-and-hounds-zeta}}

The method developed by \cite{gehan_core_2018} enables the measurement of the mean core rotation rate, which is probed by gravity modes. Hence, further methodological considerations for preparing the hares described here are focused on the gravity-dominated (g-m) dipole mixed modes, per analysis as described in \cite{gehan_core_2018}. Additionally, the operation of the hound's procedure specifically requires the most p-dominated mixed modes to be excluded from the construction of the stretched \'echelle diagram, given the \edit1{conventionally assumed inapplicability of} the technique to the most p-dominated modes \citep{mosser_period_2015, gehan_core_2018}. Nonetheless, we include them in the frequency list provided to the hound, and then consider the consistency of their properties as inferred by the hound's constraints derived from the g-dominated mixed modes.

The amplitude of a mode in the power spectrum depends strongly on $\zeta$: pure g-modes with $\zeta \to 1$ will have no intrinsic photometric amplitude, and would not be observable. Furthermore, the most g-dominated dipole mixed modes would be located in between the dipole p-modes, and therefore would be drowned out by the radial and quadrupole p-modes in the power spectrum. Thus, we must include only those modes in our hare-and-hounds exercise as might be plausible to observe even in ideal conditions.

\begin{figure}[htbp]
\centering
\includegraphics[width=.495\textwidth]{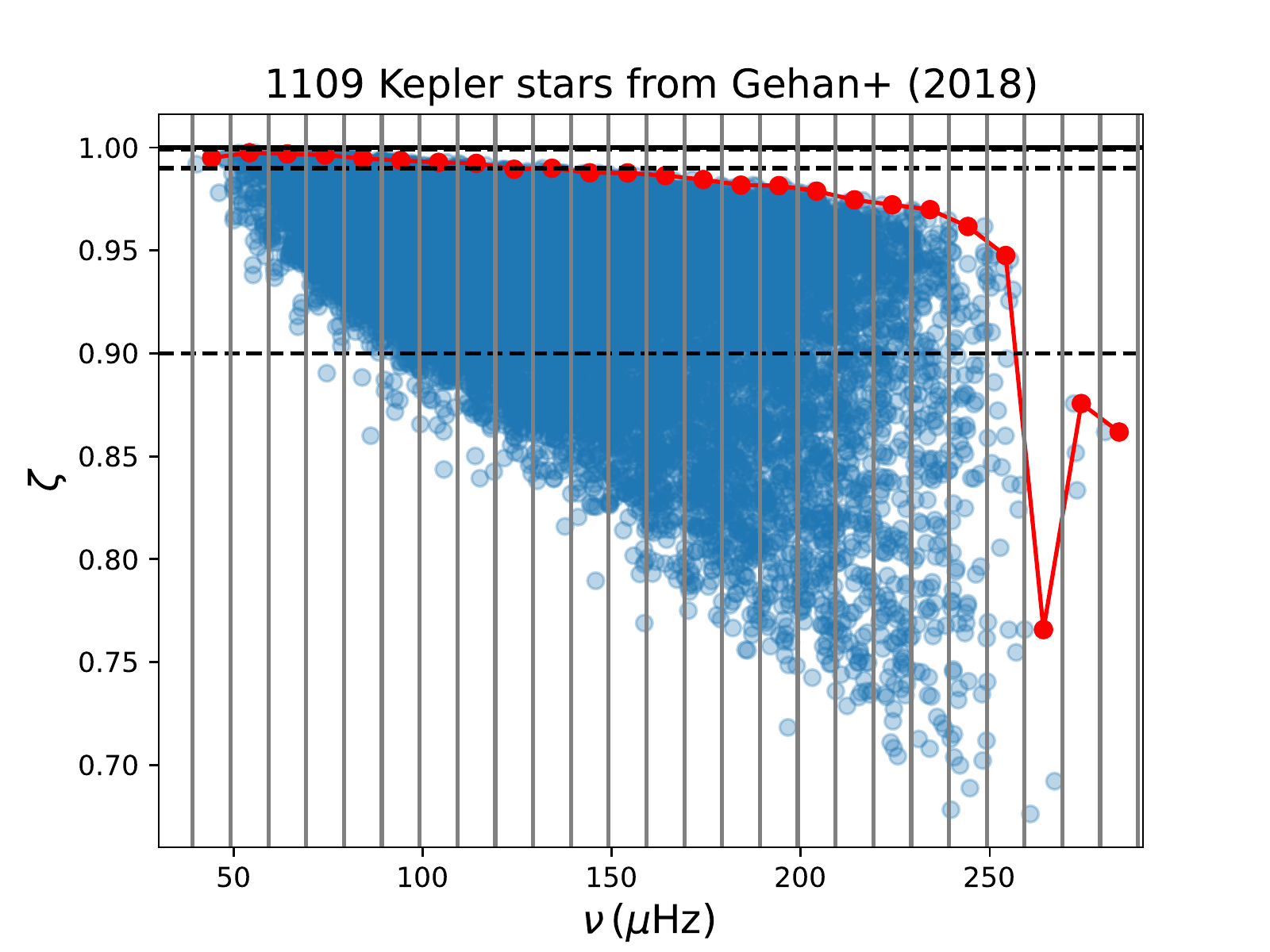}
\caption{Assumed mixing fraction $\zeta$ as inferred for 1109 \textit{Kepler} RGB stars from \cite{gehan_core_2018}, as a function of frequency. Frequency bins are delimited by vertical grey lines. The maximum $\zeta$ values within each frequency interval are represented in red (see Table~\ref{table:1}). To guide the eye, we indicate $\zeta=1$ with a horizontal black line. We also represent the values of the maximum mixing fractions used in the bottom panel of Fig. \ref{fig:rotation} by horizontal dashed lines, i.e. $\zeta\ind{max} = [1-10^{-5}, 1-10^{-4}, 1-10^{-3}, 1-10^{-2}, 1-10^{-1}]$, where the first three values are almost superimposed.} \label{fig:zeta}
\end{figure}

For this purpose, we used values for the maximal mixing fractions $\zeta$ that are consistent with observations. \cref{fig:zeta} represents the measured $\zeta$ values for all mixed modes detected over 1109 \textit{Kepler} RGB stars analyzed by \cite{gehan_core_2018}, as a function of frequency. We withheld from the hound the dipole mixed modes in our models which have $\zeta$ values above the \edit1{red envelope} displayed in \cref{fig:zeta}, for each given frequency range (see Table~\ref{table:1} in Appendix~\ref{appendix-1}). 
This serves as a representation of what would be available under ideal observing conditions. However, the most g-dominated mixed modes are also, correspondingly, the most susceptible to degradation under less ideal observing conditions than the best possible ones provide by \textit{Kepler}. To simulate the effects of progressive degradation of observing conditions, we prepared a third tranche of hares, constructed by taking hares from the evolutionary track with $M = 1.4 M_\odot$ from Set I, applying the frequency-dependent cutoff from \textit{Kepler} as above, and then further excluding modes with $\zeta$ above some threshold value. For this purpose we adopt threshold values of the  maximal mixing fractions as $\zeta\ind{max} = [1-10^{-5}, 1-10^{-4}, 1-10^{-3}, 1-10^{-2}, 1-10^{-1}]$. We will refer to these hares as Set III.

\begin{figure}[htbp]
  \centering
  \includegraphics[width=.495\textwidth]{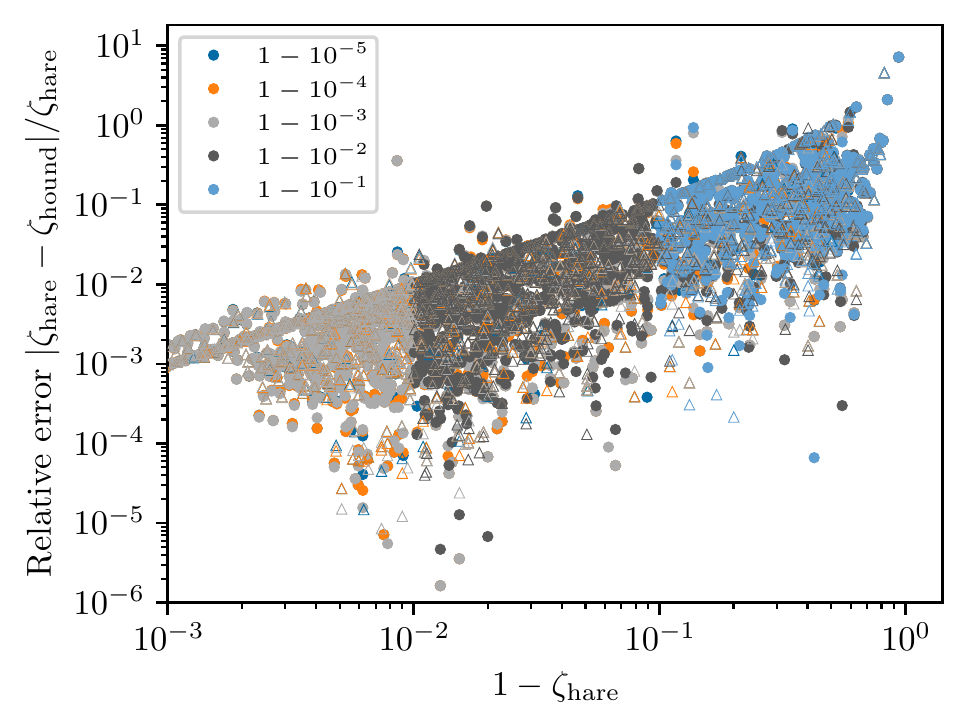}
  \caption{Relative differences between values of $\zeta$ as inferred from the hound vs. computed from the hare, shown as a function of the hare's value of $\zeta$. Points are coloured by different subsets of modes considered by the hound when performing this inference (see text for more details). Filled circles show modes accepted by the hound, while open triangles denote modes rejected by the hound as being potentially too close to the p-modes for reliable analysis.}
  \label{fig:zeta-diff}
\end{figure}

We show in \cref{fig:zeta-diff} the relative differences in the values of $\zeta$ inferred by the hound, compared to the ground-truth values available from the hare, for Set III of the hares. The overall qualitative trend is identical for the other two sets of hares, which we omit from discussion as they exhibit no visible dependence on either the mass or rotational configuration. We can see that the relative differences between the inferred and true values of $\zeta$ depend strongly on $\zeta$ itself; these determinations become increasingly unreliable as $\zeta \to 0$ --- i.e. for the most p-dominated mixed modes. Unfortunately, it is for these modes where determination of $\zeta$ is most needed, compared to the modes of the most g-like character, which to a good approximation have $\zeta \to 1$. Moreover, we see no significant trend in this pattern as more modes become available to the hypothetical observer. This suggests that this trend is intrinsic to the method by which $\zeta$ is inferred by the hound, rather than being induced by a lack of available g-dominated mixed modes with which to constrain the mixed-mode multiplet configuration. We finally note that while the algorithm employed by the hound automatically rejects modes which it deems too p-dominated to be amenable to reliable analysis, there is no clear difference in how accurately $\zeta$ is inferred between modes that have been accepted (filled circles) vs. those that have been rejected (open triangles) by the hound.

\hypertarget{hare-and-hounds-identification}{%
\subsubsection{Identification of the azimuthal order of dipole mixed modes}\label{hare-and-hounds-identification}}

\begin{figure}[htbp]
\centering
\annotate{\includegraphics[width=.495\textwidth]{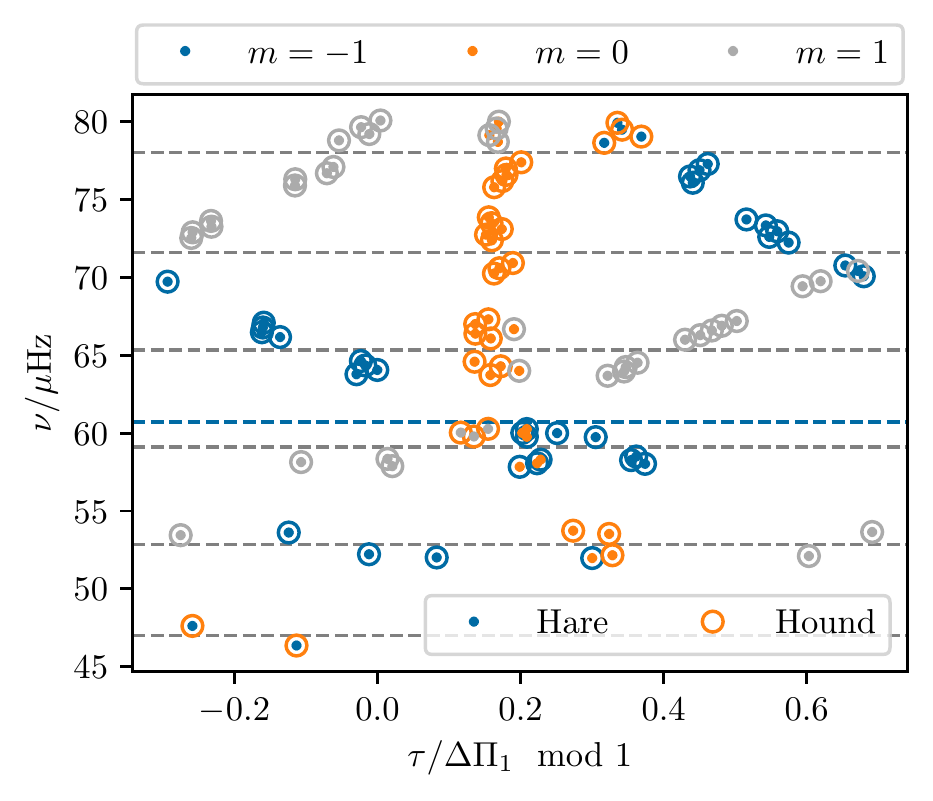}}{\node at (.21, .85){\textbf{(a)}};}
\annotate{\includegraphics[width=.495\textwidth]{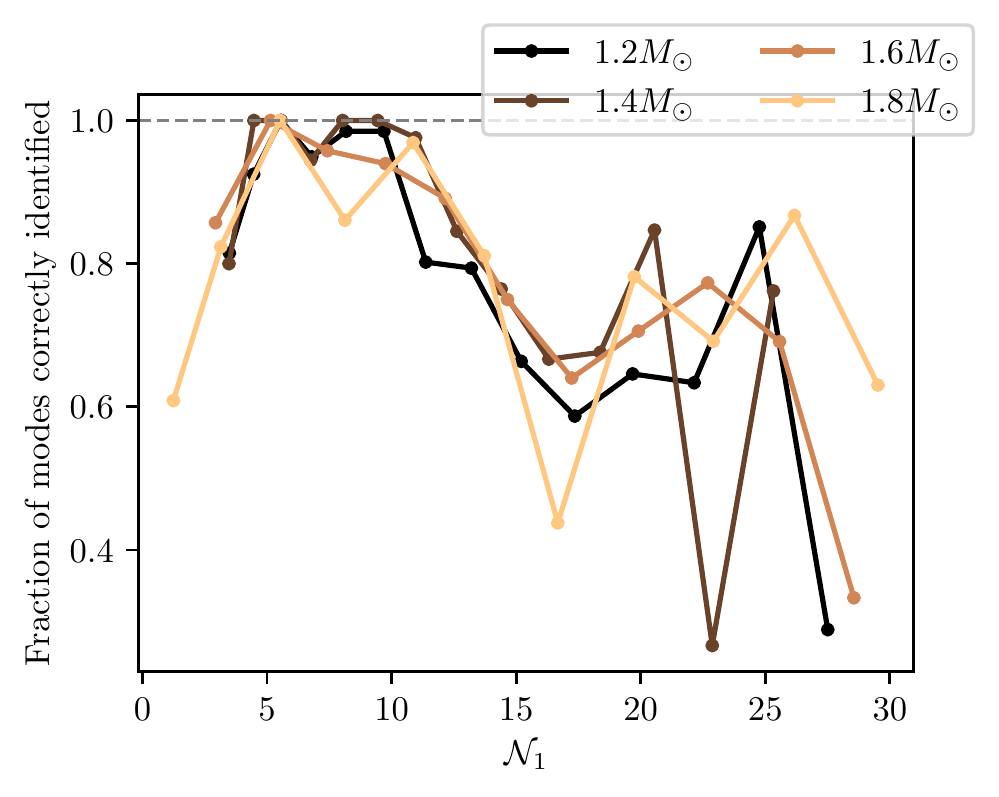}}{\node at (.21, .23){\textbf{(b)}};}
\caption{\textbf{(a)} Stretched \'echelle diagram showing misalignment in identification of multiplet components between hare and hound in Set I. This panel shows an example of misidentification for a stellar model at $M = 1.2 M_\odot, \mathcal{N}_1 \sim 25,$ and $\Omega_\text{core}/2\pi \sim 1.21\ \mu$Hz. The location of the ``fourth crossing'' \citep{gehan_2017}, where $\delta\nu_{\text{rot},g} = 2 \nu^2 \Delta\Pi_l$, is indicated with the blue dashed line. Grey dashed lines indicate the location of pure dipole p modes. \textbf{(b)} Fraction of modes correctly identified, shown over the course of evolution for different hares. Only multiplets with $\delta\nu_{\text{rot},g} < 2 \nu^2 \Delta\Pi_l$ have been included in this evaluation. The quality of the multiplet component identification can clearly be seen to degrade with increasing evolution.\label{fig:m}}
\end{figure}

We also check the consistency of the identification of the azimuthal order of dipole g-dominated mixed modes based on the method developed by \cite{gehan_core_2018}, since it can impact the measurement of the mean core rotation rate and, to a larger extent, the measurement of the inclination angle of the rotation axis \citep{gehan_automated_2021}. \cref{fig:m} (a) presents an example for a 1.2 $M\ind{\odot}$ model with a mixed-mode density $\mathcal{N}_1 = $ 25, for which the identification of the azimuthal order of dipole mixed modes differs significantly between the hare and the hound. Such misidentification essentially occurs near crossings of the rotational components of different azimuthal orders in the stretched \'echelle diagram, corresponding to local g-mode frequency spacings congruent with integer multiples of the rotational splitting, and are more susceptible to occur both when the mean core rotation rate is large (leading to larger rotational splittings), or for more evolved stars (as the g-mode spacing decreases with evolution along the RGB). This is what can be seen in \cref{fig:m} (b) for Set I, where the fraction of modes correctly identified by the hound decreases along evolution, for the four masses considered in this exercise.

\citet{gehan_core_2018,gehan_automated_2021} used this template-matching method to identify the azimuthal order of g-m mixed modes, and from these derive inclination measurements in an automated fashion. Their inclination measurements are consistent with earlier manual measurements, all of them for stars with $\mathcal{N}_1 \leq $ 15. While \cref{fig:m}(b) suggests that some of these inclination measurements ought to be revisited, we also see that $\sim$ 80 \% of the modes, or more, are correctly identified for our hares with $\mathcal{N}_1 \leq $ 15. 86.2 \% --- a substantial majority --- of the sample of \cite{gehan_automated_2021} lie within this range of evolution. Moreover, planet-hosting red giants in particular, for which such inclination measurements are of crucial astrophysical interest for understanding their dynamical architectures, live low on the red giant branch --- i.e. at lower values of $\mathcal{N}_1$ --- since significant expansion of their host stars over the course of evolution up the RGB results in planetary engulfment.

At the same time, it may be instructive to consider the various reasons behind the increasing misidentification of azimuthal orders for $N\ind{1} \gtrsim 15$. One possible reason is structural in origin, in the sense that the mixing physics adopted in generating the hares results in stellar models exhibiting sharp, localised variations of the Brunt-V\"ais\"al\"a frequency. These in turn cause buoyancy glitches: quasiperiodic modulations of the observed frequency of the g-dominated mixed modes \citep{cunha_glitches_2015, mosser_period_2015, cunha_analytical_2019}. In our models, one of these glitches corresponds to the hydrogen-burning shell, which is located at some depth in the radiative zone. This glitch produces modulations of low amplitudes, and does not significantly interact with the procedure by which the hare's modes are matched and aligned with the hound's templates. However, a second glitch is also present in our models, corresponding to the first dredge-up event on the low red giant branch, when the convective envelope reaches regions were nuclear burning previously took place \citep{salaris_dredge_up_2015}. The convective envelope then recedes as the helium core grows, leaving behind a chemical discontinuity in the gravity-mode cavity probed by the Brunt-V\"ais\"al\"a frequency. This peak in the Brunt-V\"ais\"al\"a frequency is localised much closer to the convective boundary --- i.e. at a much shallower buoyancy depth --- and results in a long-period modulation of the rotational ridges, inducing a gross curvature that is clearly visible for the $m=0$ modes in \cref{fig:m}(a). The template-matching method used by \cite{gehan_core_2018} assumes that the ridges for each $m$ are straight, which contributes to the misidentification between the hares and the hounds. \edit1{While curvature of this kind has been detected in observational mode frequencies \citep{mosser_period_2018}, it may not necessarily be uniquely attributed to buoyancy glitches \citep{deheuvels_strong_2023}, at least for first-ascent red giants \citep[as opposed to stars on the red clump;][]{vrard_evidence_2022}}. Supposing that \edit1{these glitches} are indeed astrophysically rare \citep[which would suggest the operation of additional mixing processes in actual red giants that are not present in our models; cf.][]{lindsay_overshoot_2022}, we would expect the fraction of modes correctly identified to be higher for observed red giants than for the models used in this work.

\begin{figure}[htbp]
\centering
\includegraphics[width=.495\textwidth]{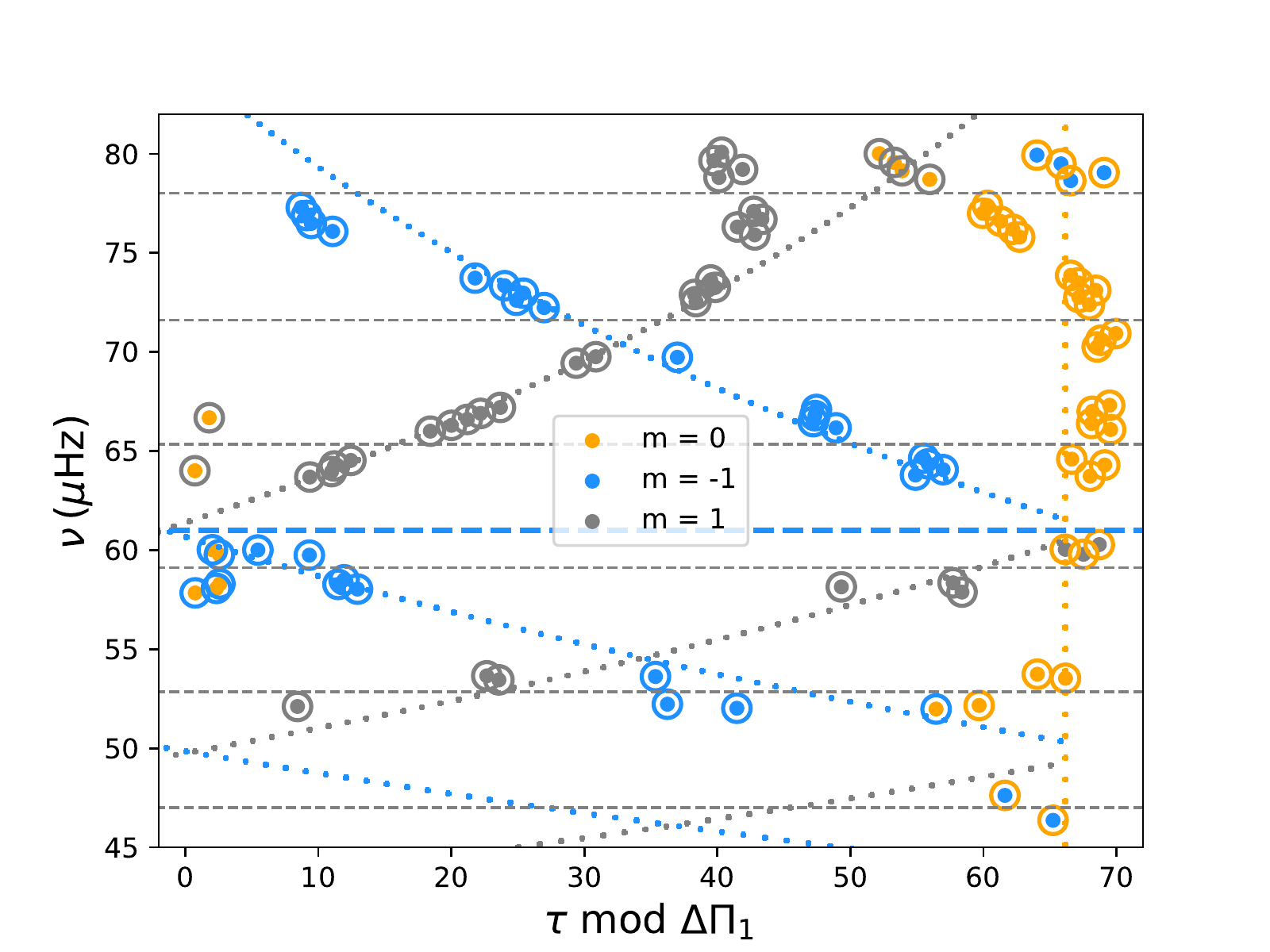}
\caption{Modes from the same stellar model as \cref{fig:m}a, but on a stretched \'echelle diagram generated by the hound --- in particular, this diagram is generated by integrating \cref{eq:ode} rather than using the closed-form expression \cref{eq:levelset}. \edit1{Otherwise, the same values of $\Delta\Pi_1$, $q$, and the p-mode frequencies are used to generate both diagrams.} Symbols and colours have the same meaning as in \cref{fig:m}a; dotted lines show the template matched by the hound against the modes. The large-scale curvature of the $m = 0$ modes can be seen to be of a different sense to that exhibited in \cref{fig:m}a. \label{fig:m-hound}}
\end{figure}

However, we have also found implementation details specific to the template-matching algorithm used in \citet{gehan_core_2018,gehan_automated_2021} to be responsible for this azimuthal-order misidentification in many other cases. While an exhaustive denumeration of failure modes is neither possible nor productive, we note some of the major ones here. In some cases where (by numerical coincidence) the rotational frequency splitting is close to some integer multiple of $\nu^2\Delta\Pi_l/2$ (i.e. half the g-mode spacing), the hound may commit an off-by-one error in identifying entire ridges on the stretched \'echelle diagram, adjusting its adopted value of $\Delta\Pi_l$ to compensate. Fortuitously, this appears not to substantially affect the value of $\Omega_\text{core}$ that the hound needs to adopt, in order to generate a template that matches the hare, even under this misidentification. In other cases, we found that numerical issues incurred in the hound's integration of \cref{eq:ode} to yield $\tau$ (which we discuss in more detail in \autoref{sec:integration}) caused the echelle diagrams obtained by the hound to be differently curved from the ground truth. For example, the curvature seen in \cref{fig:m-hound}, generated by integrating \cref{eq:ode}, is markedly different than that exhibited in \cref{fig:m}a when the closed-form expression \cref{eq:levelset} is used --- in particular, it is curved in the opposite direction. These issues resulted in template-matched misidentifications made preferentially at high and low frequencies, for essentially the same reasons as would be caused by a buoyancy glitch in the absence of these implementation issues.

\hypertarget{hare-and-hounds-rotation}{%
\subsubsection{Measurement of the mean core rotation rate}\label{hare-and-hounds-rotation}}

\begin{figure}[htbp]
\centering
\annotate{\includegraphics[width=.495\textwidth,trim=.25cm .7cm .25cm .25cm,clip]{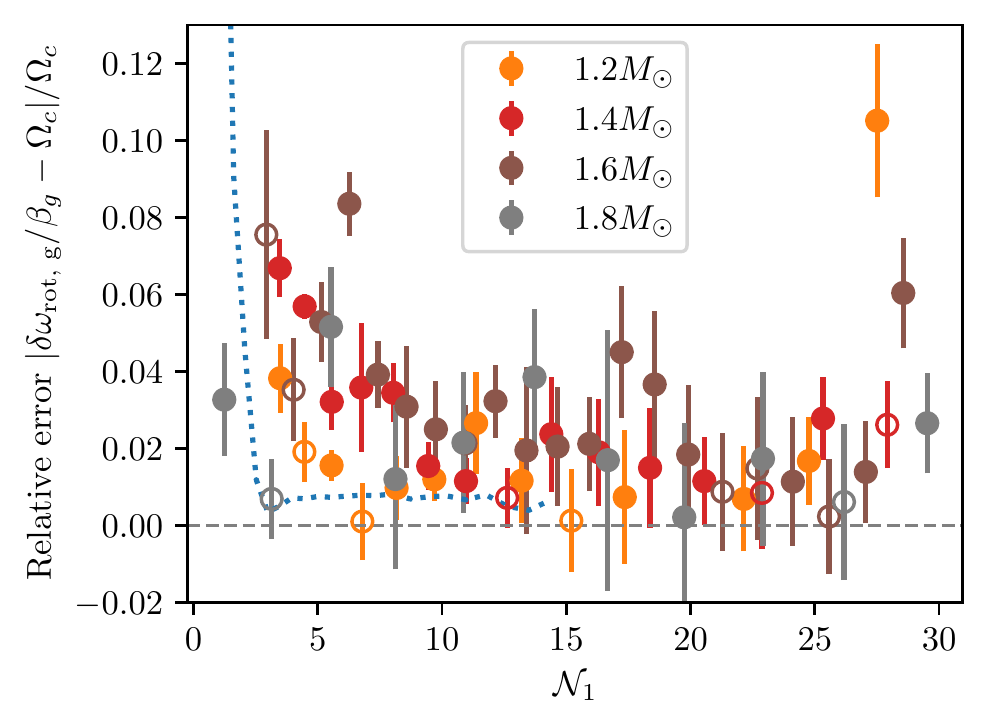}}{\node at (.35, .9){\textbf{(a)}};}
\annotate{\includegraphics[width=.495\textwidth,trim=.25cm .7cm .25cm .25cm,clip]{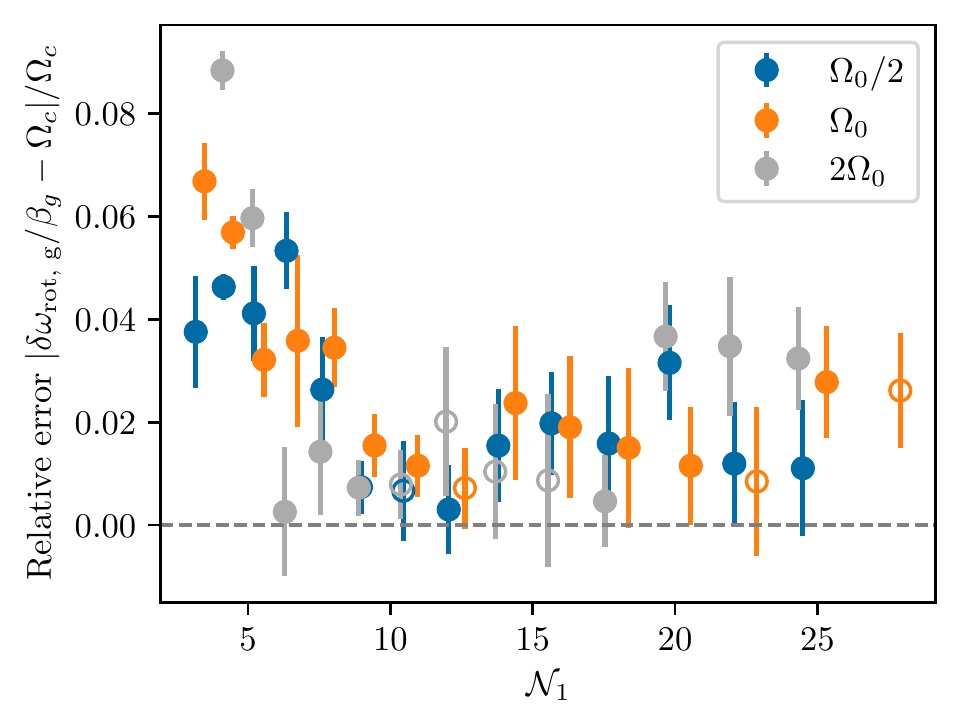}}{\node at (.21, .23){\textbf{(b)}};}
\annotate{\includegraphics[width=.495\textwidth,trim=.25cm .25cm .25cm .25cm,clip]{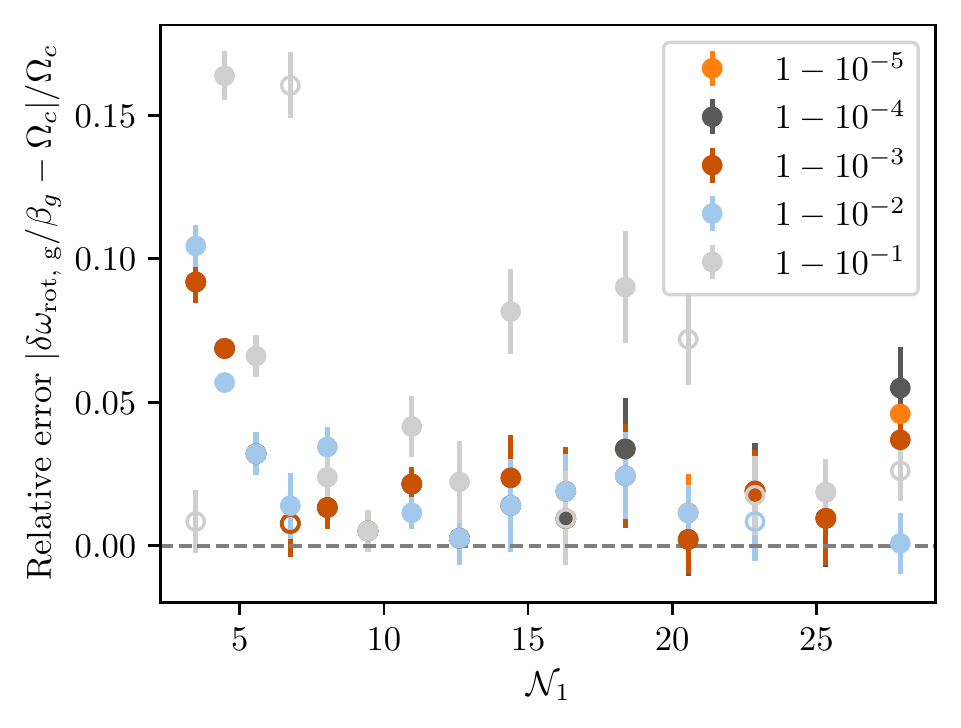}}{\node at (.21, .8){\textbf{(c)}};}
\caption{Relative errors in rotational measurements as a function of the mixed-mode density. Absolute values are shown in linear scale, with negative values indicated by open circles. The blue dotted line indicates the intrinsic error incurred by the JWKB approximation. \textbf{(a)} Impact of the stellar mass, from Set I of hares. \textbf{(b)} Impact of the mean core rotation at ZAMS, for Set II of hares. \textbf{(c)} Impact of the maximum observable mixing fraction, for Set III of hares. \label{fig:rotation}}
\end{figure}

Finally, we consider the robustness of mean core rotation rates inferred with the method of \cite{gehan_core_2018}. The relative errors in the mean core rotation measurement with respect to the ground-truth value are displayed in \cref{fig:rotation} as a function of evolution along the RGB.

In \cref{fig:rotation} (a), we show the relative differences between the g-mode rotational splitting for the hares, and the values inferred by the hounds, for Set I (examining dependences on mass). \edit1{These are expressed as relative differences between the hound's inferred values, and the} JWKB estimator of the core rotation rate \edit1{from the stellar models, which we evaluate as}
\begin{equation}
  \Omega_\text{JWKB} = {\int_{N > 0} \Omega(r) N/r\ \mathrm{d}r \over \int_{N > 0} N/r\ \mathrm{d}r}\label{eq:kern-jwkb}
\end{equation}
within the radiative core. \edit1{The rotation rates of both the hound and the JWKB estimator are assumed to match the observed splittings through \cref{eq:rotation} with $\zeta=1$, and with the g-mode sensitivity factor, $\beta_g$, assumed to take its asymptotic value of ${1 \over 2}$. However, this is known to differ from the true sensitivity factor $\beta_g$ by a small amount. Since the true value of $\beta_g$ is faithfully incorporated into the rotating mode frequencies evaluated by GYRE, but is only accounted for by the hound in such an approximate fashion, we would expect the hound's values to likewise differ from that of the JWKB estimator by such a small amount even in the event that the hound's measurements were perfectly accurate. For the $1.4M_\odot$ evolutionary track of of Set I, we thus also compare this JWKB estimator with $\beta = {1 \over 2}$ against a} ground-truth g-mode rotational splitting $(\omega_{n, m = 1} - \omega_{n, m = -1})/ 2$, averaged over the three $\gamma$-modes of the stellar model nearest to $\numax$. \edit1{We show this comparison with the blue dotted line,} up until the point where the g-mode rotational splitting becomes larger than $\numax^2 \Delta\Pi_1$, the spacing between adjacent g-modes of the same $m$ at $\numax$.

We find no significant difference between the four masses considered. The relative error in the mean core rotation measurement decreases with evolution as the number of visible mixed modes increases. This is expected since g-dominated mixed modes become more sensitive to the core along evolution, resulting in a more accurate mean core rotation measurement. The relative error in the mean core rotation measurement then tends to suddenly increase above $\mathcal{N} \sim 27$, as a result of the numerous crossings in the stretched \'echelle diagram that seriously affect the correct identification of the mixed-mode azimuthal order.

In \cref{fig:rotation} (b), we show the same relative errors for Set II of hares (examining dependences on the ZAMS rotation rate $\Omega_0$). Once again, we obtain results qualitatively similar for the different core rotation rates we have considered. The relative error in the mean core rotation measurement first decreases with evolution, reaches minimum values for $\mathcal{N} \sim [7,13]$, before taking larger values again. At large $\mathcal{N}$ values, we see that the core rotation rate slightly impacts the measured core rotation rate. Indeed, the relative error in the mean core rotation measurement significantly increases for $\mathcal{N} \gtrsim 20$ in the case of $2 \, \Omega\ind{0}$, while it is the case for $\mathcal{N} \gtrsim 25$ in the case of $\Omega\ind{0}$ and we do not see any increase in the case of $\Omega\ind{0} / 2$. As stated above, this is due to the increase in the number of crossings in the stretched \'echelle diagram along evolution. For a given mixed-mode density, a larger mean core rotation results in more crossings, leading to less accurate mean core rotation measurements. Hence, less accurate mean core rotation measurements start at lower $\mathcal{N}$ values as the mean core rotation rate increases.

In \cref{fig:rotation} (c), we show the same relative differences for Set III of hares. We obtain similar results for the different maximal mixing fractions considered, except for $\zeta\ind{max} = 1-10^{-1}$ for which we obtain significantly larger relative errors in some cases.

In summary, we find that:
\begin{itemize}
\item[(1)] the relative systematic errors on the mean core rotation rate tend to be larger than the intrinsic error from the JWKB approximation, in particular for $\mathcal{N}\ind{1} \lesssim 10$ where g-dominated mixed modes are less sensitive to the core;
\item[(2)] the mean core rotation rate as measured from the method of \cite{gehan_core_2018} tends to be systematically overestimated by at least $\sim 1-2$ \%;
\item[(3)] the minimum relative error in the mean core rotation measurement is on the order of $\sim 4$ \% or below for $\mathcal{N} \sim 7-27$;
\item[(4)] measuring accurate mean core rotation rates with a relative error of $\sim 1-2$ \% requires to have some dipole mixed modes with $\zeta > 0.90$.
\end{itemize}

We note that we infer overall quite accurate mean core rotation rates (\cref{fig:rotation}), even despite mismatches in the mode identification (\cref{fig:m}). The template matching method used by \cite{gehan_core_2018} counts the peaks that are close enough to the synthetic rotational template, and selects the configuration with the maximum number of peaks aligned with the synthetic ridges. Values of $m$ are assigned to modes post-hoc, based on the nearest synthetic ridge to each mode. As such, measuring the mean core rotation rate with the method developed by \cite{gehan_core_2018} does not require the azimuthal order of the modes to be identified at all on a mode-by-mode basis. Hence, we do not need to get a very high fraction of modes correctly identified to infer the mean core rotation rate with the method of \cite{gehan_core_2018}, as long as the synthetic stretched \'echelle diagram matches well the morphology of the observed one.

\section{Discussion}
\label{sec:discussion}

\edit1{We now briefly outline some of the implications of our findings, for various aspects of how rotational characterisation with mixed modes may be performed.}

\subsection{Numerical Integration of $\tau$}
\label{sec:integration}

We have derived a closed form expression, \cref{eq:levelset}, to the solutions of the initial value problem specified by \cref{eq:ode}. Since the latter follows from differentiating the former only when the frequency dependence of $q$ is neglected, it follows that \cref{eq:ode} cannot be used to generate diagrams with frequency-dependent $q$. \edit1{Moreover, we note that the stretched echelle diagrams shown in \cref{fig:m}a and \cref{fig:m-hound} are constructed with the same modes, same value of $\Delta\Pi$, and same value of $q$. The only difference is that in \cref{fig:m}a, the exact expression \cref{eq:levelset} is used, and in \cref{fig:m-hound}, $\tau$ is evaluated by numerical integration. Thus, the differences between the two, which are partially responsible for misidentification of $m$, must necessarily be attributed to systematics caused by such numerical integration.}

Our \edit1{results yield a simple criterion for the convergence of such numerical integration}. Specifically, the accuracy of numerical integration of an ODE of the form $y' = f(x,y)$ is determined by the relation between the integration step size to the characteristic scales in the $x$-coordinate over which $f$ varies. Our discussion of the oversampling criterion \cref{eq:sampling} required to accurately constrain the properties of $\zeta$ from observational data can therefore be adapted to also describe the step sizes required for these numerical integrations of \cref{eq:ode} to accurately reproduce the analytic solutions for the stretching function $\tau$.

\begin{figure}[htbp]
  \centering
  \includegraphics[width=.5\textwidth]{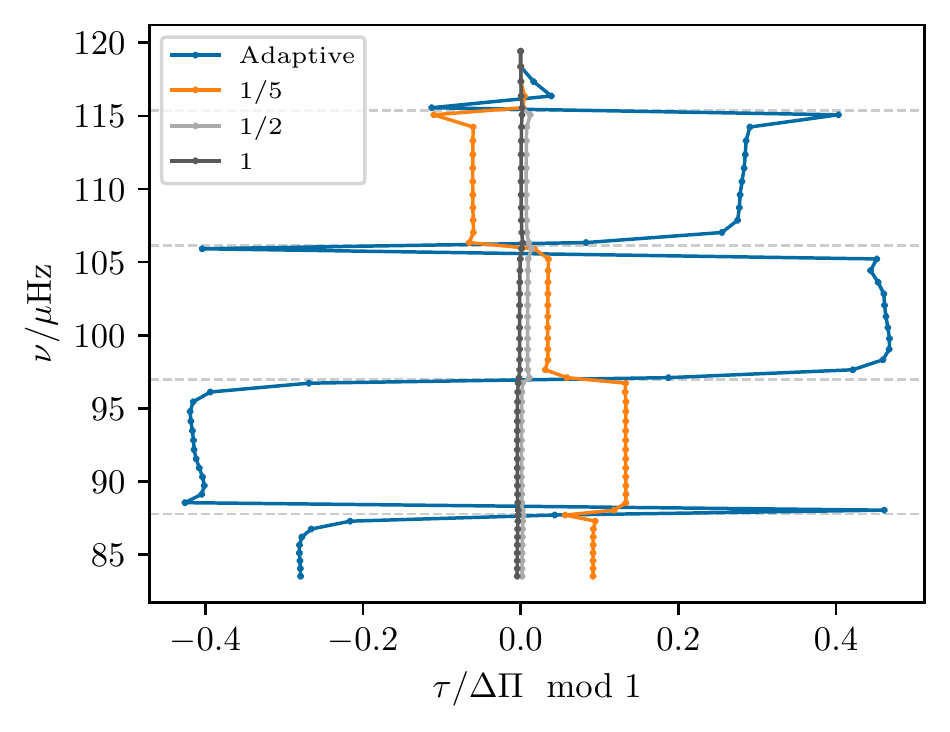}
  \caption{
  Stretched \'echelle diagrams associated with a single set of asymptotic mixed-mode frequencies, constructed with respect to stretching functions $\tau$ obtained by adaptive ODE-solving with an RK45 scheme using different maximum step sizes. Different curves indicate values of the oversampling criterion, \cref{eq:sampling-ode}, associated with the maximum step size $h_0$ passed to the numerical solver used in solving \cref{eq:ode}; smaller stepsizes yield larger sampling discriminants. The blue curve shows the results when the integrator is permitted to determine the step size entirely adaptively with no upper limit supplied. The initial condition is specified at high frequencies, where all of the solutions can be seen to agree by construction. The locations of the $\pi$ modes are indicated with the horizontal dashed lines, and can be seen to coincide with where integration error is incurred.
  }
  \label{fig:samplingode}
\end{figure}

For illustrative purposes, we compare stretched \'echelle diagrams generated using stretching functions $\tau$ computed by various numerical solutions to \cref{eq:ode}, integrated using a Runge-Kutta scheme implemented in \texttt{scipy}. For this demonstration, $\zeta_p$ is specified by the frequency-dependent $q$ (via \cref{eq:translate}), the $\pi$-mode frequencies, and idealised g-modes with the same mean $\Delta\Pi$ and $\epsilon_g$ as of a first-ascent red giant stellar model with $\numax = 100\ \mu$Hz, roughly located near the luminosity bump on the HR diagram. The integration of \cref{eq:ode} to obtain $\tau$ is performed with respect to the dimensionless period coordinate $p = P/\Delta\Pi$, with initial condition $\tau = 0$ as $p \to p_0$, where $p_0$ corresponds to the highest-frequency mode of the set of mixed modes for which we wish to generate the stretched \'echelle diagram. While this solver (RK45) is usually adaptive, we are also able to override the maximum step size permitted for use by the solver. Thus, by analogy with \cref{eq:sampling}, we may define an oversampling discriminant associated with the maximum stepsizes of the integrator, $h_0$, as
\begin{equation}
  D_\text{samp} = p^2 \gamma \Delta\Pi_1 / h_0  = p^2 q \Delta\nu \Delta\Pi_1 / h_0 \pi.\label{eq:sampling-ode}
\end{equation}
Accordingly, in addition to results from permitting the solver to adapt its stepsizes freely, we also show results obtained from setting the maximal stepsize to achieve various values of $D_\text{samp}$. Since the integrator does not accept a variable maximal stepsize, we evaluate $D_\text{samp}$ at $p = 1 / \numax \Delta\Pi_1$. We consider solutions with stepsizes chosen to yield various values of $D_\text{samp}$, indicated with different colours in \cref{fig:samplingode}.

We apply these numerical stretching functions to the corresponding idealised mixed modes associated with the specified $q$ and pure modes (generated for consistency with \cref{eq:eig}), showing the resulting stretched \'echelle diagrams in \cref{fig:samplingode}. Since the supplied mixed modes are fully consistent with the $\zeta_p$ used to generate the stretching functions, the exact solution should yield an essentially vertical line. The initial condition for integration is specified at high frequencies, and so we see that all of the numerical solutions agree there, and diverge as the integrator passes to lower frequencies. It can be seen that the fully adaptive strategy to selecting stepsizes used by the RK45 integrator yields large excursions from the correct solution in the neighbourhood of the pure p-modes. Even when smaller maximal stepsizes are enforced, it can be seen that integration error is incurred in the vicinity of the p-dominated mixed modes, with some residual variability as the integrator attempts to recover. These are reduced and eventually eliminated as $h_0$ is reduced so as to achieve $D_\text{samp} = 1$. Conversely, in order to achieve $D_\text{samp} \ge 1$, we obtain an upper limit on the permissible step sizes to be used in numerical integration of \cref{eq:ode}:
\begin{equation}
  h_0 \le {q \Delta\nu \over \pi  \nu^2 \Delta\Pi} = q \mathcal{N}_1 / \pi.
\end{equation}
Precise details regarding the numerical integration of \cref{eq:ode} have not generally been made available in most existing works employing the stretched-\'echelle-diagram construction. The integrand $f(P) \sim {1 / \zeta_p(P)}$ can be well-approximated as being mostly close to 1 except for narrow, sharp peaks near the p-mode eigenvalues \edit1{(per \cref{eq:lorentzian})}. As such, this IVP constitutes a quasiperiodic numerically stiff problem, not unlike that of integrating highly eccentric Keplerian orbits, and the details of this numerical integration are actually quite important. Beyond rotation, these issues may also interfere with the interpretation of other features of these stretched diagrams, such as potentially disrupting observational signatures of buoyancy glitches (as we saw in our discussion of \cref{fig:m-hound}). However, should the step-sizes of the integrators in these existing works be made known, this expression provides a criterion for whether or not their results ought to be reassessed. In such cases, numerical errors relating to the integration of \cref{eq:ode} may be diagnosed by comparison to the analytic expression, as we have done in our hare-and-hounds exercise.

Since this bound is proportional to $q$, we may also surmise at least that the evaluation of the stretching function by numerical integration becomes increasingly computationally expensive in the limit of weak coupling\edit1{, contraindicating doing so in the regime where we might wish to use it in conjunction with the matrix construction. Ultimately, these numerical issues may be avoided altogether} by simply using the closed-form expression for $\tau$, \cref{eq:levelset}, rather than performing any integration of \cref{eq:ode} in the first place.

\subsection{Modifications to existing asymptotic techniques}

\edit1{Template-matching techniques, as used by our hound, ordinarily require a set of mode frequencies to be measured from a power spectrum before they may be applied. However, the actual power spectrum ordinarily contains more information than just the mode frequencies: in particular, the mode linewidths and amplitudes are also known to strongly depend on the mixing fractions $\zeta$ \citep{mosser_period_2018}. Moreover, since the most g-dominated modes have the lowest amplitudes, they may not be significant enough to be reported by a standard peakbagging routine, despite being the least sensitive to misestimation of $\zeta$, and thus the most informative for the determination of $\Delta\Pi_l$ and $\Omega_\text{core}$. It would be desirable to include such constraints from these less-significant modes, however scarce, as exist in actual power spectra. Doing so would also ameliorate any ill effects which might otherwise arise when the oversampling criterion \cref{eq:sampling} is not satisfied.}

\begin{figure}[htbp]
  \centering
  \includegraphics[]{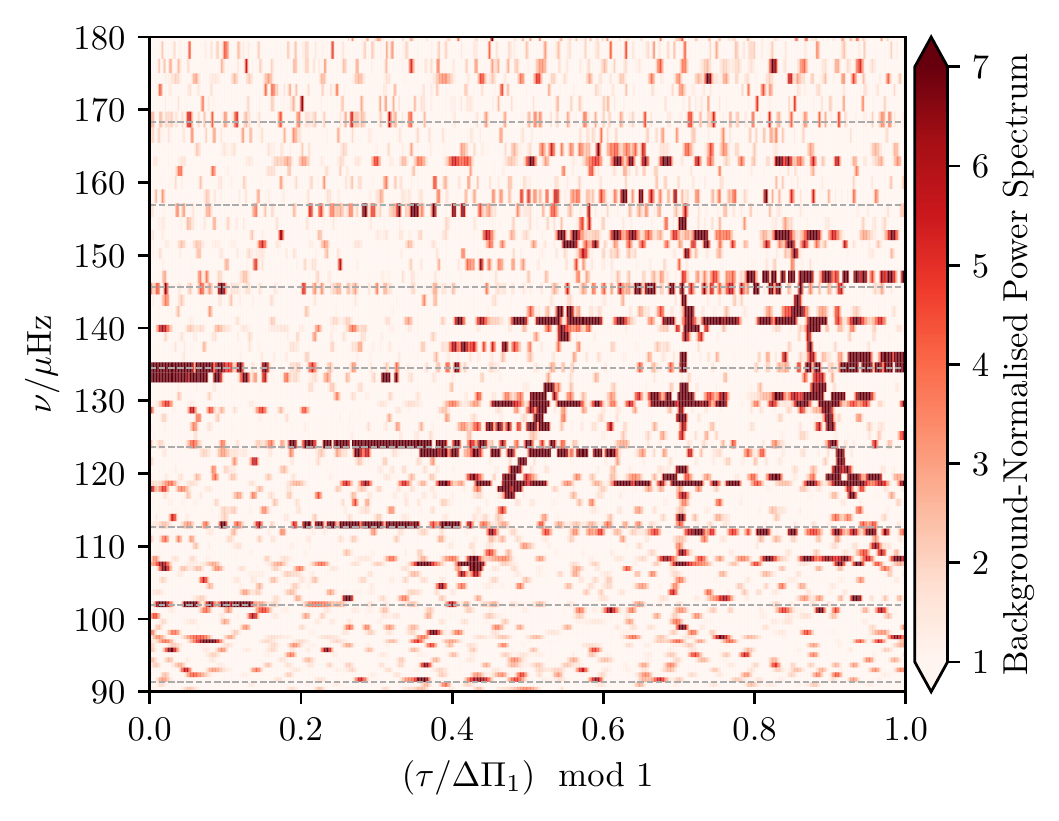}
  \caption{Illustrative stretched \'echelle power diagram of KIC 6144777, constructed with $q = 0.125$, $\Delta\Pi_1 = 79.05$ s, and with dipole p-mode frequencies derived from a fit to the even-degree modes, using the (currently in development) dipole-mode extension to \texttt{pbjam} \citep{pbjam}. The frequencies of the radial p-modes are indicated with the gray dashed lines.}
  \label{fig:powerdiagram}
\end{figure}

\edit1{Such an inclusion might be effected by template matching against stretched period-\'echelle power diagrams, as constructed in \cite{gaulme_hierarchical_2022}, rather than lists of mode frequencies. This diagram was constructed in that work by analogy with the frequency-\'echelle power diagrams appearing in p-mode asteroseismology, to provide visual diagnostics for the g-dominated mixed modes directly from the power spectrum without necessitating an initial peakbagging step. Rather than evaluating $\tau$ by ODE-solving for this purpose, we may use the closed-form expression, \cref{eq:levelset}, in this application as well. For demonstration, we show an illustrative example of such a diagram, constructed in this fashion, in \cref{fig:powerdiagram}. Aside from avoiding the above-described numerical issues and expense of ODE-solving at low $q$, doing so permits the g-mode phase offset $\epsilon_g$ to be read directly off the diagram, completely analogously to $\epsilon_p$ in frequency-echelle power diagrams; this information would be otherwise lost as a constant of integration for the ODE. Whereas \cite{gaulme_hierarchical_2022} mask out the regions containing the $\ell = 0,2$ p-modes (indicated with the dashed lines), we note that it is precisely these regions that contain the most g-dominated dipole mixed modes. In the residual power in these regions, lower-amplitude g-dominated dipole modes can be seen, which would ordinarily be filtered out by the significance testing used in many peakfinding procedures \cite[e.g.][]{corsaro_diamonds_2014}, and which would therefore be inaccessible to template matching against frequency lists. The development of techniques for template-matching directly with respect to such stretched power diagrams, as opposed to with respect to frequency lists, is likely to be a substantial undertaking, which we defer to a future work.}

\hypertarget{conclusion}{%
\subsection{Implications for rotational inversions}\label{conclusion}}

In a linear inversion procedure, it is the departures of the kernel functions from the scaled mean asymptotic kernel (specified by the JWKB wavenumber, as in \cref{eq:kern-jwkb}) which permit localisation of perturbations from a fiducial stellar model. Even in relatively less evolved red giants ($\mathcal{N}_1 \sim 5$), the kernels associated with the pure g-modes are deep into the asymptotic regime (e.g. fig. 13 of \obb), and therefore carry very little information about rotation beyond that provided by the JWKB estimator \citep[as further demonstrated by the ability of][to construct combination kernels out of mixed modes that suppress core sensitivity entirely]{ahlborn_improved_2022}. As such, any localisation provided by the mixed-mode rotational kernels will originate from the most p-dominated modes and their mixing fractions. Unfortunately, it is these mixing fractions which are least well-constrained by the asymptotic procedure. Similar considerations apply to other kinds of inversion problems, such as inversions for stellar structure and composition.

Consequently, seismic constraints on stellar rotation beyond the two-zone model, and specifically within the p-mode cavity, must draw upon additional priors on stellar evolution and structure, such as may be derived from explicit modelling. Fortunately, such modelling has to be performed anyway in order to permit the construction of appropriate inversion kernels in the first place. Given access to a sufficiently good guess to the structure of the star, the coupling strength $q$ may be then supplied a priori (from the coupling matrices via \cref{eq:translate}) rather than inferred generatively, which should in principle improve the constraints on $\zeta$, and potentially also on the notional pure p-mode rotational splittings, returned from the asymptotic procedure for the most p-dominated mixed modes. Supplying the coupling strengths in this fashion may also resolve other issues with inferring the pure p-mode frequencies themselves using the asymptotic construction (as encountered in Saunders et al., under review).

At the same time, the closed-form nature of the asymptotic parameterisation makes it far easier to operationalise, and computationally far less expensive, than the matrix decompositions employed in the algebraic approach. Moreover, \edit1{unlike the matrix procedure, numerical solutions to its eigenvalue problem are not susceptible to truncation effects, and do not require an exhaustive search of eigenvalues. Conversely,} the use of isolated p- and g-mode kernels (as we propose in \obb) precisely suppresses the sensitivity of the inversion procedure to the inferred mixing fractions of the observed modes, thereby avoiding some \edit1{of the} shortcomings \edit1{of the asymptotic procedure that we have illustrated with our hare-and-hounds exercise.}

\section{Conclusion}

Our ultimate objective in this series of papers is to place constraints on the internal rotation of stars, beyond the presently-used one- or two-zone model of radial differential rotation. In \obb, we found that the asymptotic formulation underlying the stretched-\'echelle-diagram construction had to be modified to treat each $m$ separately in order to permit this; in this work we assess various other features of techniques employing this construction. \edit1{Since certain liberties taken in the $\zeta$-function construction are thought to limit its applicability to the most p-dominated mixed modes, we first produce a derivation that fills some of these conceptual deficiencies, and so show that in fact the current construction of the stretching function ought to be applicable to them as well --- i.e. we first produce a "steel man" of the asymptotic construction that is as rigorous as we can make it. Given this exposition, we find that even within this fortified construction, a configuration of mixed modes must still satisfy a certain criterion, \cref{eq:sampling,eq:sampling2}, to provide enough information to constrain the true function $\tau$ describing it, and that most of this information comes from the p-dominated mixed modes. As a bonus, our geometric formulation yields a closed-form expression for the stretching function $\tau$ underlying stretched \'echelle diagrams,} permitting it to be evaluated in closed form via \cref{eq:levelset}, rather than by solving the initial value problem specified by \cref{eq:ode}, as is usually done.

\edit1{From stellar modelling we find that red giant models in the range of $\mathcal{N}$ spanned by the observational sample appear not to satisfy this information criterion, indicating that characterisations of the mode coupling strengths of the avoided crossings, and thus mixing fractions for the most p-dominated mixed modes, are potentially misestimated --- and systematically so --- for this observational sample. We perform a hare-and-hounds exercise to determine how this might affect measurements of not just g-mode core rotation (i.e. the object of astrophysical interest for existing studies), but also of $\zeta$ and of mode identifications (of importance to the next steps in this inversion-oriented plan of work), as recovered by an existing observational pipeline using this asymptotic parameterisation --- in particular, the template-matching algorithm developed by and used in \citet{gehan_core_2018,gehan_automated_2021}, which relies on the stretched-\'echelle construction. The hare-and-hounds exercise shows that robust recovery of the former, but not of the latter (particularly for the most p-dominated modes), is possible when the information criterion is not satisfied.} In particular, the identification of multiplet components decreases in accuracy as the coupling strength of the mixed-mode system weakens, and the mixing fraction $\zeta$ is identified least accurately for the most p-dominated mixed modes. However, some of these shortcomings stem from other methodological features of the algorithm that has played the role of the hound in this exercise, rather than being limitations of the asymptotic parameterisation per se\edit1{, or issues stemming from the mischaracterisation of $q$}.

\edit1{In parallel to this, we compare the algebraic construction (which allows the generation of inversion kernels with desirable localisation properties; see \obb) to the asymptotic one, so that in cases where the asymptotic construction is usable, we can immediately adapt its results for use with inversions. In \obb{} we already show that envelope rotation is not accessible without significantly modifying current applications of the asymptotic construction; as above, we show here that other ingredients required for inversions (namely $\zeta$ for the most p-dominated modes, and the multiplet component identifications in general) are also methodologically inaccessible as well. To complete the comparison,} we have derived an explicit translation between quantities appearing in the asymptotic construction, and in the nonasymptotic, matrix-based approach to mixed-mode coupling. Specifically, we find that the asymptotic eigenvalue equation can be obtained as the weak-coupling limit of the nonasymptotic eigenvalue equation, and that strict equivalence between the two formulations requires further constraints to be placed on the frequency dependence of the strength of the coupling between the two mode cavities. We interpret this as indicating that the nonasymptotic description is most well-approximated by the asymptotic one in the limit of weak coupling.

\edit1{Since we have demonstrated that the two formulations become increasingly interoperable in the limit of weak coupling} --- particularly after the turnoff in $q$ with increasing $\mathcal{N}$ as seen in \cref{fig:disc}a --- \edit1{the use of features from each may remedy limitations of the other, particularly for the most evolved red giants. Such synergy may} prove useful in also addressing an outstanding methodological gap in operationalising the algebraic approach for inversions, being that the notional pure p- and g-mode rotational splittings must be inferred from the observed mixed modes. We defer an investigation of this problem, and of the localisation properties of these isolated kernels, to future work.

We thank S. Basu, B. Mosser, J. van Saders, and M. Cunha for constructive feedback on this work. \edit1{We also thank the anonymous referee for their unusually voluminous criticism and feedback, which have substantially improved the quality of the finished paper}. JO acknowledges support from NASA through the NASA Hubble Fellowship grant HST-HF2-51517.001 awarded by the Space Telescope Science Institute, which is operated by the Association of Universities for Research in Astronomy, Incorporated, under NASA contract NAS5-26555. CG was supported by Max Planck Society (Max Planck Gesellschaft) grant “Preparations for PLATO Science” M.FE.A.Aero 0011. We make available our implementation of the analytic expressions derived in this paper\edit1{, and a helper routine to construct stretched \'echelle power diagrams,} through a Python package hosted at \url{https://gitlab.com/darthoctopus/zeta}.

\appendix

\section{Verification of an identity involving $\zeta$ in the algebraic construction}
\label{app:verify}

We proceed by direct differentiation, almost exactly following the derivation of eq. B12 in \citet{ong_semianalytic_2020}. We will first consider both mixed-mode eigenvalues at once, and specialise to the g-dominated mixed mode later. On the right-hand-side of \cref{eq:matrixspacing}, we have
\begin{equation}
  \begin{aligned}
  {\partial \omega_\pm \over \partial \omega_g} &= {\omega_g \over \omega_\pm} {\partial \omega_\pm^2 \over \partial \omega_g^2} = {\omega_g \over \omega_\pm} \left({1 \over 2} \pm {\omega_g^2 - \omega_p^2 \over 2\sqrt{\left(\omega_p^2 - \omega_g^2\right)^2 + 4\alpha^2}}\right) \\
  &= \pm {\omega_g \over \omega_\pm} \left( {\omega_g^2 - \omega_p^2 \pm \sqrt{\left(\omega_p^2 - \omega_g^2\right)^2 + 4\alpha^2} \over 2 \sqrt{\left(\omega_p^2 - \omega_g^2\right)^2 + 4\alpha^2}}\right) \\
  &= {\omega_g \over \omega_\pm} \left( {\omega_\pm^2 - \omega_p^2 \over  \pm\sqrt{\left(\omega_p^2 - \omega_g^2\right)^2 + 4\alpha^2}}\right).
  \end{aligned}
\end{equation}
For the left-hand-side of \cref{eq:matrixspacing}, we first note that \cref{eq:u} implies
\begin{equation}
  u_\pm^2 = {\omega^2_\pm - \omega_p^2 \over \omega^2_\pm - \omega_g^2}; 1 + u_\pm^2 =  {2\omega^2_\pm - \omega_p^2 - \omega_g^2 \over \omega^2_\pm - \omega_g^2} = {\pm\sqrt{\left(\omega_p^2 - \omega_g^2\right)^2 + 4\alpha^2} \over \omega^2_\pm - \omega_g^2}.
\end{equation}
Thus, we obtain
\begin{equation}
  \zeta_\pm = {u_\pm^2 \over 1 + u_\pm^2} = {\omega^2_\pm - \omega_p^2 \over \pm\sqrt{\left(\omega_p^2 - \omega_g^2\right)^2 + 4\alpha^2}} = {\omega_\pm \over \omega_g} {\partial \omega_\pm \over \partial \omega_g}.
\end{equation}
For the more g-dominated of the two solutions, $\omega_g \sim \omega_\pm$, thereby yielding the desired result. However, for a pair of modes close to resonance, this is a good approximation for both modes in the limit of weak coupling.

\section{Maximum measured mixing fractions}\label{appendix-1}

In Table~\ref{table:1} we indicate the maximum mixing fractions $\zeta\ind{max}$ that we measure for each given frequency range in \cref{fig:zeta}. We report the maxima of modes binned in intervals 10 $\mu$Hz wide; the frequency values shown here are the midpoints of these intervals.

\begin{table}[h!]
\caption{Maximum mixing fraction $\zeta$ measured for 1109 \textit{Kepler} RGB stars, as a function of frequency, as displayed in red in \cref{fig:zeta}.} 
\label{table:1}
\centering
\begin{tabular}{cc}
\toprule
$\nu \, (\mu$Hz) & $\zeta\ind{max}$\\
\midrule
44  & 0.995 \\
54  & 0.997 \\
64  & 0.997 \\
74  & 0.996 \\
84  & 0.995 \\
94  & 0.994 \\
104 & 0.993 \\
114 & 0.992 \\
124 & 0.989 \\
134 & 0.990 \\
144 & 0.988 \\
154 & 0.988 \\
164 & 0.986 \\
174 & 0.984 \\
184 & 0.982 \\
194 & 0.981 \\
204 & 0.979 \\
214 & 0.975 \\
224 & 0.972 \\
234 & 0.970 \\
244 & 0.962 \\
254 & 0.948 \\
264 & 0.766 \\
274 & 0.876 \\
284 & 0.862 \\
\bottomrule
\end{tabular}
\end{table}

\bibliography{biblio.bib}
\end{document}